\documentclass[11pt]{article}
\usepackage{amsmath,amssymb}
\newcommand{\de}{\mathrm{d}}
\newcommand{\ee}{\mathrm{e}}
\newcommand{\im}{\mathrm{i}}
\newcommand{\ft}[2]{{\textstyle\frac{#1}{#2}}}

\begin{document}
\begin{titlepage}
\begin{center}
\hfill LMU-ASC 02/06\\
\hfill ITP-UU-06/01 \\
\hfill SPIN-06/01   \\
\hfill AEI-2006-002  \\
\hfill LTH 687     \\
\hfill {\tt hep-th/0601108}\\
\vskip 6mm

{\Large \textbf{Black hole partition functions and duality }}
\vskip 8mm

\textbf{G.L.~Cardoso$^{a}$, B. de Wit$^b$, J.~K\"appeli$^{c}$
       and  T. Mohaupt$^{d}$}

\vskip 4mm
$^a${\em Arnold Sommerfeld Center for Theoretical Physics\\
Department f\"ur Physik,
Ludwig-Maximilians-Universit\"at M\"unchen, Munich, Germany}\\
{\tt gcardoso@theorie.physik.uni-muenchen.de}\\[1mm]
$^b${\em Institute for Theoretical Physics} and {\em Spinoza
  Institute,\\ Utrecht University, Utrecht, The Netherlands}\\
{\tt  B.deWit@phys.uu.nl} \\[1mm]
$^c${\em Max-Planck-Institut f\"ur Gravitationsphysik,
  Albert-Einstein-Institut, Potsdam, Germany}\\
{\tt kaeppeli@aei.mpg.de}\\[1mm]
$^d${\em Department of Mathematical Sciences, University of Liverpool,
Liverpool L69 3BX, U.K.}\\
{\tt Thomas.Mohaupt@liverpool.ac.uk}
\end{center}
\vskip .2in
\begin{center} {\bf ABSTRACT } \end{center}
\begin{quotation}\noindent
  The macroscopic entropy and the attractor equations for BPS black
  holes in four-dimensional $N=2$ supergravity theories follow from a
  variational principle for a certain `entropy function'. We present
  this function in the presence of $R^2$-interactions and
  non-holomorphic corrections.  The variational principle identifies
  the entropy as a Legendre transform and this motivates the
  definition of various partition functions corresponding to different
  ensembles and a hierarchy of corresponding duality invariant inverse
  Laplace integral representations for the microscopic degeneracies.\\
  Whenever the microscopic degeneracies are known the partition
  functions can be evaluated directly. This is the case for $N=4$
  heterotic CHL black holes, where we demonstrate that the partition
  functions are consistent with the results obtained on the
  macroscopic side for black holes that have a non-vanishing classical
  area. In this way we confirm the presence of a measure in the
  duality invariant inverse Laplace integrals. Most, but not all, of
  these results are obtained in the context of semiclassical
  approximations. For black holes whose area vanishes
  classically, there remain discrepancies at the semiclassical level
  and beyond, the nature of which is not fully understood at present.
\end{quotation}

\vfill
\end{titlepage}

\eject
\section{Introduction}
\setcounter{equation}{0}
The degeneracy of BPS states for certain wrapped brane or string
configurations, which can be identified with extremal black holes,
defines a statistical or microscopic entropy. This statistical entropy
can be successfully compared to the macroscopic entropy for the
extremal black holes that arise as supersymmetric solutions of the
effective field theory associated with the corresponding string
compactification \cite{Strominger:1996sh}. Initially, this comparison
made use of the Bekenstein-Hawking area law for black hole entropy.
More refined calculations \cite{Maldacena:1997de,Vafa:1997gr} of the
asymptotic degeneracy of microstates revealed that there are
corrections to the area law.  Subsequently, it was demonstrated how
higher-order derivative couplings based on chiral $N=2$ superspace
densities in the effective action account for a successful agreement
with the microscopic results \cite{LopesCardoso:1998wt}. A necessary
ingredient in this work is provided by Wald's definition of black hole
entropy based on a Noether surface charge \cite{Wald:1993nt}, which
ensures the validity of the first law of black hole mechanics.  This
definition enabled the derivation of a general thermodynamic or
macroscopic entropy formula for the $N=2$ supergravity theories
discussed above. Here a crucial role is played by the fixed-point
behavior: at the black hole horizon supersymmetry enhancement forces
some of the fields, and in particular the moduli, to fixed values
determined by the electric and magnetic charges $q$ and $p$ carried by
the black hole. This attractor phenomenon persists for the
supergravity theories with higher-derivative interactions
\cite{LopesCardoso:2000qm}. The macroscopic entropy is therefore a
function solely of the black hole charges.  Adopting this generalized
notion of black hole entropy and the black hole attractor behaviour,
agreement of the macroscopic entropy has been established with the
known asymptotic microstate degeneracies to subleading order in the
limit of large charges.

More recently these refinements have led to further insights and
conjectures. In \cite{Ooguri:2004zv} it was observed that the
thermodynamic entropy formula \cite{LopesCardoso:1998wt} including the
full series of higher-derivative corrections can be rewritten as the
Legendre transform of a real function $\mathcal{F}(p,\phi)$ with
respect to the electrostatic potentials $\phi$ defined at the black
hole horizon. The electric charges are retrieved by $q= \partial{\cal
  F}/\partial\phi$.  Remarkably, the `free energy'
$\mathcal{F}(p,\phi)$ obtained in this way from the thermodynamic
entropy is related to the topological string partition function
$Z_{\mathrm{top}}(p,\phi)$ by the simple relation (we use the
normalizations of this paper)
\begin{equation}
  \label{eq:toplogical}
   \mathrm{e}^{\pi \,\mathcal{F}(p,\phi)} =\vert
   Z_{\mathrm{top}}(p,\phi)  \vert^2  \,.
\end{equation}
According to the conjecture of \cite{Ooguri:2004zv}, the function
$\mathcal{F}$ on the left-hand side should be interpreted as the free
energy associated with a 
black hole partition function defined in terms of the microscopic 
degeneracies $d(p,q)$, which for given charges $p$ and $q$ define the
microcanonical partition function. In view of the above relation the
black hole ensemble relevant for the comparison to topological strings
is the one where the magnetic charges $p$ and the electrostatic
potentials $\phi$ are held fixed. With respect to the magnetic charges
one is therefore dealing with a microcanonical ensemble, while the
quantized electric charges are replaced by the continuous electrostatic
potentials $\phi$ when passing to a canonical ensemble by a Laplace
transformation \cite{Ooguri:2004zv},
\begin{equation}
  \label{eq:partdegen}
  Z(p,\phi) = \sum_{\{ q\}} d(p,q) \, \ee^{\pi\,q\cdot \phi}\,.
\end{equation}
The conjecture is thus that the mixed microcanonical/canonical black
hole partition function is given by 
\begin{equation}
  \label{eq:origOSVconj}
 Z(p,\phi) \approx  \mathrm{e}^{\pi\,\mathcal{F}(p,\phi)}\,,
\end{equation}
which, through (\ref{eq:toplogical}), is related to the topological
string. 
 
As the effective action formed the starting point for the above
conjecture, it is clear that there exists in any case an indirect
relation with the topological string. The genus-$g$ partition
functions of the topological string \cite{Bershadsky:1993ta} are known
to be related to certain higher-derivative interactions in an $N=2$
supersymmetric string effective action. The holomorphic anomaly
associated with these partition functions is related to non-Wilsonian
terms in the effective action associated with the propagation of
massless states \cite{Antoniadis:1994}. The crucial question is
therefore to understand what the implications are of this conjecture
beyond its connection to the effective action. Further work in
that direction can be found in
\cite{Vafa:2004qa,Aganagic:2004js,Caporaso:2005np,Aganagic:2005wn},
where the conjecture was tested for the case of non-compact
Calabi-Yau spaces. Other work concerns the question of how the
background dependence related to the holomorphic anomaly equations
and how the wave function interpretation of the topological string
partition functions are encoded in the black hole partition
functions.  Interesting progress in this direction can be found in
\cite{Gerasimov:2004yx,Verlinde:2004ck, Ooguri:2005vr}.

Viewing $Z(p,\phi)$ as a holomorphic function in $\phi$, the relation
(\ref{eq:origOSVconj}) can be used to express the microscopic black
hole degeneracies as an inverse Laplace transform,
\begin{equation}
  \label{eq:OSV-int}
  d(p,q) = \int \de\phi \, Z(p,\phi) \, \ee^{-\pi \,q\cdot\phi}
  \approx\int   {\mathrm d}\phi \;
  \mathrm{e}^{\pi[\mathcal{F}(p,\phi) - q\cdot\phi]}   \,. 
\end{equation} 
In the limit of large charges the result of the integral is expected
to be equal to the exponent of the Legendre transform of
$\pi\mathcal{F}$, which is, by definition, the macroscopic entropy
that formed the starting point. The question is then whether
(\ref{eq:OSV-int}) captures certain of the subleading corrections
encoded in the microscopic degeneracies $d(p,q)$. Various results have
been obtained to this extent, mostly for the case of 1/2-BPS black
holes in $N=4$ string theory
\cite{Dabholkar:2004yr,Dabholkar:2005x,Sen:2005ch,Dabholkar:2005dt}.
There are of course questions regarding the convergence of
(\ref{eq:partdegen}) and the required periodicity of
$\exp[\pi\mathcal{F}(p,\phi)]$ under imaginary shifts in $\phi$. The
latter is, conversely, related to the necessity of having to specify
integration contours for the complex $\phi$-integrations when
extracting black hole degeneracies from $\exp{[\pi\mathcal{F}(p,\phi)]}$
using (\ref{eq:OSV-int}).

While many questions seem to be primarily related to technical
complications and must be discussed in a case-by-case fashion, the
issue of covariance with respect to electric-magnetic duality
transformations can be addressed in fairly broad generality. It forms
the main subject of this paper.  The status of electric-magnetic
duality covariance of the original proposal (\ref{eq:origOSVconj}) is
at first somewhat obscured by the fact that one is working with a
mixed canonical/microcanonical black hole ensemble and is therefore
not treating the electric and magnetic charges on equal footing. At
first sight it is therefore not obvious what electric-magnetic duality
covariance implies at the level of (\ref{eq:origOSVconj}).
Furthermore, the black hole degeneracies obtained through
(\ref{eq:OSV-int}) should be consistent with duality symmetries such
as S- or T-duality.

In this paper we start from the fully canonical black hole partition
function depending on the electro- and magnetostatic potentials $\phi$ and
$\chi$, which are conjugate to the quantized electric and magnetic
charges $q$ and $p$,
\begin{equation}
  \label{eq:fullycanon}
  Z(\phi, \chi) = \sum_{\{p,q \}} d(p,q) 
  \, \mathrm{e}^{\pi[q\cdot\phi-p\cdot\chi]}\,.
\end{equation}
Is there a function $\mathrm{e}^{2\pi \mathcal{H}(\chi,\phi)}$ that is
(at least in semiclassical approximation) associated to $Z(\phi,
\chi)$ in analogy to the original conjecture?  If such a function
exists, what is its relation to the function 
$\mathrm{e}^{\pi\mathcal{F}(p,\phi)}$ of (\ref{eq:origOSVconj})? The
answers to these questions turn out to be intimately related to the
existence of a variational principle for black hole entropy. The
associated entropy function naturally accommodates both the
higher-order derivative terms and certain non-holomorphic
interactions. The strategy of this paper consists in uncovering this
variational principle and thereby identifying
$\mathrm{e}^{2\pi\mathcal{H}(\chi,\phi)}$. Then, using that
$Z(p,\phi)$ is related to $Z(\phi,\chi)$ by an inverse Laplace
transform with respect to $\chi$, subleading corrections to the
proposal (\ref{eq:origOSVconj}) are derived at the semiclassical
level. These corrections appear as measure factors when retrieving
black hole degeneracies as in (\ref{eq:OSV-int}) and implement the
requirement of covariance under
electric-magnetic duality transformations.\footnote{
  \label{footnote1}
  The consequences of this approach have been discussed by the authors
  at recent conferences. These include the `Workshop on gravitational
  aspects of string theory' at the Fields Institute (Toronto, May
  2005) and `Strings 2005' (Toronto, July 2005;
  http://www.fields.utoronto.ca/audio/05-06/strings/wit/index.html).
  See also \cite{Kappeli:2005nm, deWit:2005ya,Mohaupt:2005jd}.}

Our approach can be tested in cases where the microscopic degeneracies
are known. This is the case for heterotic black holes in $N=4$
supersymmetric string theory, where for the so-called CHL models the
exact degeneracies of $1/4$\,- and $1/2$\,-BPS states are known. In
the limit of large charges, the $1/4$-BPS states correspond to regular
dyonic black holes carrying both electric and magnetic charges, whose
area is much bigger than the string scale. Hence these black holes
are called `large'. The $1/2$-BPS black holes are either
electrically or magnetically charged and their area is of order of the
string scale. At the two-derivative level the effective action leads
to a vanishing area. These black holes are called `small'.

The exact dyon degeneracies are encoded in certain automorphic
functions, from which both the asymptotic degeneracies and the
dominant contributions to the partition function can be extracted. In
this paper we demonstrate that in this way the microscopic data indeed
yield the macroscopic results and, in particular, confirm the presence
of the measure factors in integrals such as (\ref{eq:OSV-int})
and generalizations thereof. While these results are extremely
satisfactory, we should stress that at present there is no
evidence that the correspondence can be extended beyond the
semiclassical level. Nevertheless the agreement in the dyonic case is
impressive as it involves non-perturbative terms in the string
coupling. 

The microstates of the $1/2$-BPS black holes are the perturbative
string states and their microcanonical partition function is therefore
known. Here an analogous comparison with the macroscopic results
turns out to be rather intricate and agreement is found at leading
order only. In our opinion there is at this moment no 
satisfactory way to fully account for the relation between microscopic
and macroscopic descriptions of the small black holes at the
semiclassical level and beyond, in spite of the fact that partial
successes have been reported. In this paper we note that the
semiclassical description seems to depend sensitively on the
higher-derivative corrections and even on the presence of the
non-holomorphic corrections, so that reliable calculations are
difficult. Beyond this observation we have not many clues as to what
is actually responsible for the rather general lack of agreement,
which is in such a sharp contrast to the situation encountered for the
large black holes.

The outline of this paper is as follows. In
section~\ref{sec:variational} the variational principle that underlies
both the black hole attractor mechanism and black hole entropy is
introduced. It is explained how to incorporate both the
higher-derivative interactions as well as non-holomorphic
interactions. In subsection~\ref{sec:hesse} the variational principle
is reformulated in terms of real coordinates. These real coordinates
will correspond to the electro- and magnetostatic potentials measured
at the black hole horizon. In section~\ref{sec:partial-L-duality} the
variational principle is worked out for the case when only a
restricted set of variables is varied while the others are kept fixed
at their attractor values. One thereby recovers, for example, the
original observations of \cite{Ooguri:2004zv}.  The various entropy
functions obtained in this way are worked out for $N=4$ models and the
role of duality transformations and of non-holomorphic corrections is
explained.  In section~\ref{sec:partition-functions} we identify the
various free energies obtained from the variational principles with
partition sums over corresponding ensembles involving the microscopic
degeneracies. We generally prove that for cases where the
semiclassical approximation is appropriate, the macroscopic and
microscopic descriptions for large black holes are in agreement.
In particular the effects of the proposed measure factors are
elucidated for generic $N=2$ models and subsequently worked out for
the $N=4$ examples under consideration. We describe the discrepancies
that arise for small black holes. In section~\ref{sec:heterotic-bh} 
the partition functions of dyonic $1/4$-BPS and of $1/2$-BPS black
holes are considered for the CHL models. We review the agreement with
the macroscopic results of the asymptotic degeneracies and present a
direct calculation of the mixed partition function. The latter is in
agreement with the measure factors derived earlier on the basis of the
macroscopic results. However, the agreements only pertain to the large
black holes. The troublesome features noted before for the small black
holes persist here as well.

\section{Macroscopic entropy as a Legendre transform}
\label{sec:variational}
\setcounter{equation}{0}
Lagrangians for $N=2$ supergravity coupled to vector supermultiplets
that depend at most quadratically on space-time derivatives of the
fields, are encoded in a homogeneous holomorphic function $F(X)$ of
second degree. Here the complex $X^I$ are related to the vector multiplet
scalar fields (henceforth called moduli), up to an overall
identification by a complex space-time dependent factor. The
Lagrangian does not depend on this function but only on its
derivatives. The index $I=0,1,\ldots,n$ labels all the vector fields,
including the graviphoton, so that the matter fields comprise $n$
vector supermultiplets. The $(2n+2)$-component vector $(X^I,F_I)$,
whose components are sometimes called `periods' in view of their
connection with the periods of the holomorphic three-form of a
Calabi-Yau three-fold, play a central role. Here $F_I$ is defined by
$F_I= \partial F/\partial X^I$. Under electric/magnetic duality
transformations these periods rotate under elements of
$\mathrm{USp}(n+1,n+1)$.  It is possible to describe $(X^I,F_I)$ as
the holomorphic sections of a line bundle, but this is not needed
below.

The function $F(X)$, and therefore the $F_I(X)$, can be modified by
extra (holomorphic) terms associated with the so-called Weyl
supermultiplet of supergravity. These modifications give rise to
additional interactions involving higher space-time derivatives; the
most conspicuous coupling is the one proportional to the square of
the Weyl tensor. Furthermore the effective action will contain
non-local interactions whose generic form has, so far, not been
fully determined.  These non-local interactions induce
non-holomorphic terms in the $F_I(X)$ which are needed for realizing
the invariance of the full effective action under symmetries that
are not respected by the Wilsonian effective action.  The latter
action is based on holomorphic quantities and it describes the
effect of integrating out massive degrees of freedom.  Both these
holomorphic and non-holomorphic modifications will play an important
role in this paper.

Supersymmetric (BPS) black hole solutions exhibit the so-called
attractor phenomenon
\cite{Ferrara:1995ih,Strominger:1996kf,Ferrara:1996dd}, which implies
that at the horizon the scalar moduli take values that are fixed in
terms of the electric and magnetic charges of the black hole.
Henceforth these charges will be denoted by $q_I$ and $p^I$,
respectively.  Because the entropy is based on the horizon properties
of the various fields, the attractor mechanism ensures that the
macroscopic entropy can be expressed entirely in terms of the charges.
The attractor equations originate from the fact that the BPS solutions
exhibit supersymmetry enhancement at the horizon. Globally the BPS
solution has residual $N=1$ supersymmetry, but locally, at the horizon
and at spatial infinity, the solution exhibits full $N=2$
supersymmetry.  Although the attractor mechanism was originally
discovered in the supersymmetric context, it has been known for a
while \cite{Gibbons,Ferrara:1997tw} that it can also occur in a more
general non-supersymmetric context. Recent studies discussing various
aspects of this have appeared in
\cite{Sen:2005wa,Goldstein:2005hq,Sen:2005iz,Kallosh:2005ax,
  Tripathy:2005qp,Giryavets:2005nf,Prester:2005qs,
  Goldstein:2005rr,Alishahiha}.  
For instance, in \cite{Sen:2005wa,Sen:2005iz} it was shown that the
attractor mechanism also holds in the context of non-supersymmetric
extremal black hole solutions in covariant higher-derivative gravity
theories, provided one makes certain assumptions on the horizon
geometry. 

Since $(X^I,F_I)$ and $(p^I,q_I)$ transform identically under
electric/magnetic duality, it is not surprising that the attractor
equations define a linear relation between the period vector
$(X^I,F_I)$, its complex conjugate, and the charge vector $(p^I,q_I)$.
However, in view of the fact that the $X^I$ are defined up to a
complex rescaling it is clear that there should be a certain
normalization factor whose behaviour under the rescalings is such that
the resulting expression is invariant. This normalization can be
absorbed into the definition of $X^I$ and leads to the quantities
$Y^I$ and $F_I = \partial F(Y)/\partial Y^I$ which are no longer
subject to these rescalings, although the function $F(Y)$ inherits, of
course, the scaling properties of the original function $F(X)$
\cite{Behrndt:1996jn}.  Performing the same rescaling to the square of the
lowest component of the Weyl multiplet (this is an auxiliary tensor
field that is usually called the graviphoton `field strength'), we
obtain an extra complex scalar denoted by $\Upsilon$. On the horizon
the values of $Y^I$, $F_I$ and $\Upsilon$ are fixed by the attractor
equations,
\begin{equation}
  \label{eq:attractor}
  Y^I-\bar Y^I = \mathrm{i} p^I\,,\qquad
  F_I(Y,\Upsilon) - \bar F_I(\bar Y,\bar\Upsilon) = \mathrm{i} q_I\,,
  \qquad \Upsilon= -
  64\,.
\end{equation}
Here we introduced a possible holomorphic dependence of the function
$F$ on the Weyl multiplet field $\Upsilon$ which will induce $R^2$
terms and other higher-derivative terms in the Wilsonian action.
Supersymmetry requires the function $F(Y,\Upsilon)$ to be homogeneous
of second degree,
\begin{equation}
  \label{eq:homogeneous-F}
  F(\lambda Y,\lambda^2 \Upsilon) = \lambda^2\, F( Y,\Upsilon)\,.
\end{equation}
For the moment we ignore the issue of possible non-holomorphic
corrections and first proceed in a holomorphic setting.

As stated in the introduction, the macroscopic black hole entropy
follows from a variational principle. To see this, define the `entropy
function',
\begin{equation}
  \label{eq:Sigma-simple}
  \Sigma(Y,\bar Y,p,q) =  \mathcal{F}(Y,\bar Y,\Upsilon,\bar\Upsilon)
  - q_I   (Y^I+\bar Y^I)   + p^I (F_I+\bar F_I)  \;,
\end{equation}
where $p^I$ and $q_I$ couple to the corresponding magneto- and 
electrostatic
potentials at the horizon (cf. \cite{LopesCardoso:2000qm}) in a way
that is consistent with electric/magnetic duality. The quantity
$\mathcal{F}(Y,\bar Y,\Upsilon,\bar\Upsilon)$ will be denoted as the
`free energy' for reasons that will become clear. For the case at hand
$\mathcal{F}$ is given by
\begin{equation}
  \label{eq:free-energy-phase}
  \mathcal{F}(Y,\bar Y,\Upsilon,\bar\Upsilon)= - \im \left( {\bar Y}^I
  F_I - Y^I {\bar F}_I
  \right) - 2\mathrm{i} \left( \Upsilon F_\Upsilon - \bar \Upsilon
  \bar F_{\Upsilon}\right)\,,
\end{equation}
where $F_\Upsilon= \partial F/\partial\Upsilon$. Also this expression
is compatible with electric/magnetic duality \cite{deWit:1996ix}. Varying
the entropy function $\Sigma$ with respect to the $Y^I$, while keeping
the charges and $\Upsilon$ fixed, yields the result,
\begin{equation}
  \label{eq:Sigma-variation-1}
 \delta \Sigma = \mathrm{i} ( {Y}^J - {\bar Y}^I - i p^I)\, \delta
( F_{I} + \bar F_I)  - \mathrm{i} ( F_I - {\bar F}_I - i q_I)\,
 \delta (Y^I+ \bar Y^I)  \;.
\end{equation}
Here we made use of the homogeneity of the function $F(Y)$.  Under the
mild assumption that the matrix $N_{IJ} = 2\,\mathrm{Im}\, F_{IJ}$ is
non-degenerate, it thus follows that stationary points of
$\Sigma$ satisfy the attractor equations.\footnote{
  In the absence of $\Upsilon$-dependent terms this variational
  principle was first proposed in \cite{Behrndt:1996jn}. Observe that
  it pertains specifically to black holes that exhibit supersymmetry
  enhancement at the horizon.}  
Moreover, at the stationary point, we have $q_IY^I-p^IF_I = 
-\mathrm{i} \left( {\bar Y}^I F_I - Y^I {\bar F}_I \right)$. The
macroscopic entropy is equal to the entropy function taken at the
attractor point. This implies that the macroscopic entropy is the
Legendre transform of the free energy $\mathcal{F}$.  An explicit
calculation yields the entropy formula obtained in
\cite{LopesCardoso:1998wt},
\begin{equation}
  \label{eq:W-entropy}
  {\cal S}_{\rm macro}(p,q) = \pi \, \Sigma\Big\vert_{\rm attractor}
    = \pi \Big[\vert Z\vert^2 - 256\,
  {\rm Im}\, F_\Upsilon \Big ]_{\Upsilon=-64} \,,
\end{equation}
where $\vert Z\vert^2=p^I F_I - q_I Y^I$. Here the first term
represents a quarter of the horizon area (in Planck units) so that the
second term defines the deviation from the Bekenstein-Hawking area
law. In view of the homogeneity properties and the fact that
$\Upsilon$ takes a fixed value (namely the attractor value
$\Upsilon=-64$), the second term will be subleading in the limit of
large charges. Note, however, that also the area will contain
subleading terms, as it will also depend on $\Upsilon$. In the absence
of $\Upsilon$-dependent terms, the homogeneity of the function $F(Y)$
implies that the area scales quadratically with the charges. The
$\Upsilon$-dependent terms define subleading corrections to this
result. 

In the introduction we already mentioned that there exist black hole
solutions whose horizon vanishes in the classical approximation
\cite{Sen:1994eb,Sen:1995in}. In that case the leading contribution to
the macroscopic entropy originates entirely from $R^2$-interactions
and scales only linearly with the charges. For example, this happens
for black holes corresponding to certain perturbative heterotic $N=4$
string states \cite{Dabholkar:1989jt}.  These black holes are called
`small' black holes in view of their vanishing classical area, while
the generic ones are called `large' black holes.  We will adopt this
terminology throughout this paper.

We now extend the above results to incorporate the non-holomorphic
corrections. As it turns out \cite{Cardoso:2004xf}, this extension is effected
by changing the function $F(Y,\Upsilon)$ to
$F(Y,\Upsilon)+2\mathrm{i}\, \Omega(Y,\bar Y,\Upsilon,\bar\Upsilon)$,
where $\Omega$ is real and homogeneous of second degree. When $\Omega$
equals the imaginary part of a holomorphic function of $Y$ and
$\Upsilon$, so that $\Omega$ is harmonic, we can always absorb the
holomorphic part into $F(Y,\Upsilon)$ and drop the anti-holomorphic
part.  Alternatively, this implies that all the $\Upsilon$-dependent
terms can always be absorbed into $\Omega$ and this observation will
be exploited later on. The shift of the function $F$ induces the
following changes in the derivatives $F_I$, $\bar F_I$ and
$F_\Upsilon$,
\begin{equation}
  \label{eq:shift}
  F_I \to F_I + 2 \mathrm{i} \, \Omega_I  \;, \qquad
  F_{\Upsilon} \to  F_{\Upsilon} +
  2 \mathrm{i} \, \Omega_{\Upsilon} \;,
\end{equation}
where $\Omega_I = \partial \Omega/\partial Y^I$, $\Omega_{\bar I} =
\partial \Omega/\partial \bar Y^I$ and $\Omega_{\Upsilon} = \partial
\Omega/\partial \Upsilon$. Note that for holomorphic functions we do
not use different subscripts ($I$ and $\bar I$, or $\Upsilon$ and
$\bar\Upsilon$, respectively) to distinguish holomorphic and
anti-holomorphic derivatives. The homogeneity implies,
\begin{eqnarray}
  \label{eq:homog}
  2 \,F -Y^I F_I &=& 2\,\Upsilon F_\Upsilon \,, \nonumber\\
  2 \,\Omega - Y^I \,\Omega_I - {\bar Y}^I \,\Omega_{\bar I}
  &=& 2 \,\Upsilon \,\Omega_{\Upsilon} + 2 {\bar \Upsilon}\,
  \Omega_{\bar \Upsilon} \;.
\end{eqnarray}
Substituting (\ref{eq:shift}) in the free energy $\mathcal{F}$, one
obtains the following modified expression,
\begin{eqnarray}
  \label{eq:free-energy-Omega}
  \mathcal{F}(Y,\bar Y,\Upsilon,\bar\Upsilon) &=& - \im \left( {\bar Y}^I
  F_I - Y^I {\bar F}_I
  \right) - 2\mathrm{i} \left( \Upsilon F_\Upsilon - \bar \Upsilon
  \bar F_{\Upsilon}\right) \nonumber \\
  &&
  + 4 \,\Omega - 2 (Y^I - {\bar Y}^I)(\Omega_I - \Omega_{\bar I}) \;.
\end{eqnarray}
Here we made use of the second equation of (\ref{eq:homog}). However,
also the corresponding expression of the entropy function will be
modified,
\begin{equation}
  \label{eq:Sigma-Omega}
\Sigma(Y,\bar Y,p,q) =  \mathcal{F}(Y,\bar Y,\Upsilon,\bar\Upsilon)
  - q_I   (Y^I+\bar Y^I)   + p^I (F_I+\bar F_I +
    2\im(\Omega_I-\Omega_{\bar I}))  \;.
\end{equation}
With this definition the variation of the entropy function induced by
$\delta Y^I$ and $\delta \bar Y^I$ reads,
\begin{eqnarray}
  \label{eq:var1-Sigma-Omega}
  \delta \Sigma &=& \mathrm{i} (Y^I - {\bar Y}^I -i p^I) \,\delta (F_I+\bar
  F_I + 2\mathrm{i}(\Omega_I-\Omega_{\bar I})) \nonumber\\
  &&{}
  - \mathrm{i} (F_I-\bar F_I +2\mathrm{i} (\Omega_I + \Omega_{\bar
    I})-\im q_I) \,\delta(Y^I + \bar Y^I) \,,
\end{eqnarray}
which is a straightforward generalization of
(\ref{eq:Sigma-variation-1}). Its form confirms that $\Omega$ can be
absorbed into the holomorphic $F_I$ when $\Omega$ is harmonic.
Stationary points of the modified entropy function thus satisfy the
following attractor equations,
\begin{equation}
  \label{eq:nonholoattr}
  Y^I - {\bar Y}^I = \im p^I \,, \qquad \hat F_I - \bar{\hat F}_I =
  \im q_I \,,
\end{equation}
where here and henceforth we use the notation $\hat F_I$ to indicate
the modification by $\Omega$,
\begin{equation}
  \label{eq:hat-F-I}
  \hat F_I = F_I +2\im\,\Omega_I\,.
\end{equation}
This leads to the definition of a modified period vector whose
components consist of the $Y^I$ and the $\hat F_I$, where the latter
will, in general, no longer be holomorphic. This description for
incorporating non-holomorphic corrections is in accord with the
approach used in \cite{LopesCardoso:1999ur}, where a function $\Omega$
was constructed for heterotic black holes by insisting that the
(modified) periods transform consistently under S-duality. From the
effective action point of view, the non-holomorphic contributions to
$\Omega$ originate from the non-local invariants that must be included
in the effective action. From the topological string point of view,
these terms are related to the holomorphic anomaly which is due to a
non-holomorphic dependence of the genus-$g$ partition functions on
the background \cite{Bershadsky:1993ta,Witten:1993ed}.

The issue of electric/magnetic duality is subtle in the presence of
non-holomorphic corrections. We discuss it in subsection
\ref{sec:T-S-duality} when analyzing T- and S-duality for $N=4$
heterotic black holes. S-duality requires the presence of
non-holomorphic terms, which leads to an entropy function that is
invariant under both T- and S-duality. To obtain the entropy one
evaluates the entropy function at the attractor point. The result is
precisely equal to (\ref{eq:W-entropy}) upon changing the function $F$
into $F+2\im\,\Omega$. Hence the entropy is the Legendre
transform of the free energy (\ref{eq:free-energy-Omega}).

In the following subsection~\ref{sec:hesse} we consider the variational
principle and the corresponding Legendre transform in terms of real
variables corresponding to the electrostatic and magnetostatic
potentials. This shows that the macroscopic entropy is in fact a
Legendre transform of the so-called Hesse potential (or its
appropriate extension).

\subsection{A real basis and the Hesse potential}
\label{sec:hesse}
In this subsection we present a reformulation of the variational
principle in terms of real variables. This allows us to find an
interpretation of the full Legendre transform of the entropy in the
context of special geometry.  In special geometry one usually employs
complex variables, but in the context of BPS solutions, it is the real
and imaginary parts of the symplectic vector $(Y^I,F_I)$ that play a
role. Namely, the imaginary part is subject to the attractor
equations, whereas the real part defines the electrostatic and
magnetostatic potentials \cite{LopesCardoso:2000qm}. Therefore it is
no surprise that the form of the variational formulae
(\ref{eq:Sigma-simple}) and (\ref{eq:Sigma-variation-1}) suggests a
formulation in terms of $2(n+1)$ real variables equal to the
potentials, rather than in terms of the real and imaginary parts of
the $n+1$ complex variables $Y^I$. The conversion between the two sets
of coordinates is well-defined whenever $N_{IJ} = 2\,\mathrm{Im}\,
F_{IJ}$ is non-degenerate. The discussion of special geometry in terms
of the real coordinates can be found in
\cite{Freed:1997dp,Alekseevsky}.  It turns out that the prepotential
of special geometry has a real counterpart \cite{Hitchin}, the Hesse
potential, which is related to the imaginary part of the holomorphic
prepotential by a Legendre transform \cite{Cortes:2001qd}.  The
distinction between complex and real polarizations also played a role
in the interpretation of the topological partition function as a wave
function on moduli space \cite{Verlinde:2004ck}. The purpose of this
subsection is to exhibit the variational principle in the context of
these real variables and to show that the black hole entropy is just
the Legendre transform of the (generalized) Hesse potential.

At this point we include the $R^2$-corrections encoded by $\Upsilon$,
but ignore the non-holomorphic correction, which will be dealt with
later.  The independent complex fields are $(Y^I, \Upsilon)$, and
associated with them is the the holomorphic function $F(Y,\Upsilon)$,
which is homogeneous of second degree (\ref{eq:homog}). We start by
decomposing $Y^I$ and $F_I$ into their real and imaginary parts,
\begin{equation}
  \label{eq:u-y}
  Y^I = x^I + \im u^I \;,\qquad F_I = y_I + \im v_I \;,
\end{equation}
where $F_I = F_I(Y,\Upsilon)$.  The real parametrization is obtained
by taking $(x^I,y_I, \Upsilon, {\bar \Upsilon})$ instead of $(Y^I,
{\bar Y}^I, \Upsilon, {\bar \Upsilon})$ as the independent variables.
Although $\Upsilon$ is a spectator, note that the inversion of $y_I =
y_I(x,u,\Upsilon,{\bar \Upsilon})$ gives ${\rm Im}\; Y^I =
u^I(x,y,\Upsilon, {\bar \Upsilon})$. To compare partial derivatives in
the two parametrizations, we need (we refrain from explicitly
indicating the $\Upsilon$-dependence),
\begin{eqnarray}
  \label{eq:derivative-u,y}
  \frac{\partial}{\partial x^I}\Big\vert_u &=& \frac{\partial}{\partial
  x^I}\Big\vert_y + \frac{\partial y_J(x,u)}{\partial x^I} \,
  \frac{\partial}{\partial y_J}\Big\vert_x\;, \nonumber\\
  \frac{\partial}{\partial u^I}\Big\vert_x &=& \frac{\partial
  y_J(x,u)}{\partial u^I} \, \frac{\partial}{\partial
  y_J}\Big\vert_x\;, \nonumber\\
  \frac{\partial}{\partial \Upsilon}\Big\vert_{x,u} &=&
  \frac{\partial}{\partial \Upsilon}\Big\vert_{x,y} + \frac{\partial
  y_I(x,u)}{\partial \Upsilon} \, \frac{\partial}{\partial
  y_I}\Big\vert_x\;.
\end{eqnarray}
The homogeneity will be preserved under the reparametrization in view
of the fact that $y(x,u)$ is a homogeneous function of first
degree. This is reflected in the equality,
\begin{eqnarray}
  \label{eq:homogeneity-u,y}
  &&
  x^I\,\frac{\partial}{\partial x^I}\Big\vert_u+
  u^I\,\frac{\partial}{\partial u^I}\Big\vert_x +
  2\, \Upsilon\,\frac{\partial}{\partial \Upsilon}\Big\vert_{x,u} +
  2\, \bar\Upsilon\,\frac{\partial}{\partial
  \bar\Upsilon}\Big\vert_{x,u} \nonumber\\
&&{}=
  x^I\,\frac{\partial}{\partial x^I}\Big\vert_y+
  y_I\,\frac{\partial}{\partial y_I}\Big\vert_x +
  2\, \Upsilon\,\frac{\partial}{\partial \Upsilon}\Big\vert_{x,y} +
  2\, \bar\Upsilon\,\frac{\partial}{\partial
  \bar\Upsilon}\Big\vert_{x,y}\;.
\end{eqnarray}
It is straightforward to write down the inverse of
(\ref{eq:derivative-u,y}),
\begin{eqnarray}
  \label{eq:derivative-y,u}
  \frac{\partial}{\partial x^I}\Big\vert_y &=& \frac{\partial}{\partial
  x^I}\Big\vert_u + \frac{\partial u^J(x,y)}{\partial x^I} \,
  \frac{\partial}{\partial u^J}\Big\vert_x\;, \nonumber\\
  \frac{\partial}{\partial y_I}\Big\vert_x &=& \frac{\partial
  u^J(x,y)}{\partial y_I} \, \frac{\partial}{\partial
  u^J}\Big\vert_x\;, \nonumber\\
  \frac{\partial}{\partial \Upsilon}\Big\vert_{x,y} &=&
  \frac{\partial}{\partial \Upsilon}\Big\vert_{x,u} + \frac{\partial
  u^I(x,y)}{\partial \Upsilon} \, \frac{\partial}{\partial
  u^I}\Big\vert_x\;.
\end{eqnarray}
Combining (\ref{eq:derivative-u,y}) with (\ref{eq:derivative-y,u})
enables one to find explicit expressions for the derivatives of
$y(x,u)$ and $u(x,y)$. One can easily verify that the
reparametrization is not well defined when $\det(N_{IJ})=0$.

Since the Hesse potential occurring in special geometry is twice the
Legendre transform of the imaginary part of the prepotential with
respect to $u^I=\mbox{Im}\,Y^I$, we define the generalized Hesse
potential by
\begin{equation}
  \label{eq:GenHesseP}
  \mathcal{H}(x,y,\Upsilon, {\bar \Upsilon}) = 2 \; {\rm Im}\,F(x+\im
  u,\Upsilon,\bar\Upsilon)  - 2 \,  y_I \,u^I \;,
\end{equation}
which is a homogeneous function of second degree. With the help
of (\ref{eq:homog}) we find
\begin{equation}
  \label{eq:GenHesseP2}
  \mathcal{H}(x,y,\Upsilon, {\bar \Upsilon}) = - \ft{1}2 \im( {\bar
  Y}^I F_I - {\bar F}_I Y^I )
  - \im (\Upsilon F_\Upsilon - {\bar \Upsilon} {\bar F}_{{\bar
    \Upsilon}}) \;,
\end{equation}
which is just proportional to the free energy defined in
(\ref{eq:free-energy-phase}). However, while the term proportional to
$\mathrm{Im}\, \Upsilon F_\Upsilon$ in (\ref{eq:free-energy-phase})
was introduced in order to obtain the correct attractor equations,
this term is now a consequence of the natural definition
(\ref{eq:GenHesseP}), as we see explicitly in (\ref{eq:GenHesseP2}).  It
is gratifying to see that the corresponding variational principle thus
has an interpretation in terms of special geometry. The entropy
function (\ref{eq:Sigma-simple}) is now replaced by
\begin{equation}
  \label{eq:RealSigma}
  \Sigma(x,y,p,q) = 2\,\mathcal{H}(x,y,\Upsilon,\bar\Upsilon) - 2 \,q_I
  x^I + 2\,p^I y_I \;.
\end{equation}
Indeed, extremization of $\Sigma$ with respect to $(x^I,y_I)$ yields
\begin{equation}
  \label{eq:ExtremEqs}
  \frac{\partial \mathcal{H}}{\partial x^I} = q_I \;,\qquad
  \frac{\partial \mathcal{H}}{\partial y_I} = - p^I  \;.
\end{equation}
Using the relations (\ref{eq:derivative-y,u}) it is straightforward to
show that the extremization equations (\ref{eq:ExtremEqs}) are just
the attractor equations (\ref{eq:attractor}), written in terms of the
new variables $(x^I,y_I)$. Substituting (\ref{eq:ExtremEqs}) into
$\Sigma$ one can verify that the Legendre transform of $\mathcal{H}$ is
proportional to the entropy (\ref{eq:W-entropy}),
\begin{eqnarray}
  \label{eq:entropy-hesse}
  {\cal S}_{\rm{macro}}(p,q) &=& 2 \pi\Big[\mathcal{H} - x^I
  \frac{\partial \mathcal{H}}{\partial x^I}
  - y_I \frac{\partial \mathcal{H}}{\partial y_I} \Big]_{\rm{attractor}}
\nonumber \\
&=&
  2 \pi\Big[- \mathcal{H} +  2\, \Upsilon \frac{\partial
    \mathcal{H}}{\partial \Upsilon}+  2\, \bar\Upsilon \frac{\partial
    \mathcal{H}}{\partial \bar\Upsilon} \Big]_{\rm attractor}\,,
\end{eqnarray}
where we used the homogeneity of $\mathcal{H}$. This expression
coincides with (\ref{eq:W-entropy}) as
$\partial\mathcal{H}/\partial\Upsilon\vert_{x,y}= -\im F_\Upsilon$.

Let us finally also include the non-holomorphic corrections.  Using
that $\mathcal{H}$ and $\Omega$ are homogeneous functions of second
degree we find from (\ref{eq:shift}) and (\ref{eq:GenHesseP2}) that
adding the non-holomorphic corrections amounts to the replacement
\begin{equation}
  \label{eq:hesse-nonholo}
  \mathcal{H} \rightarrow \hat{\mathcal{H}} = \mathcal{H} + 2 \Omega -
  (Y^I - {\bar Y}^I) (\Omega_I  - \Omega_{\bar I}) \;.
\end{equation}
Since $\mathcal{H}$ is the Legendre transform of $2 \,\mathrm{Im} \,F$, we
see that $\hat{\mathcal{H}}$ is the Legendre transform of $2\,\mathrm{Im}\,
F(x+\im u,\Upsilon,\bar\Upsilon) +2\, \Omega(x,u,\Upsilon,\bar\Upsilon)$,
which is proportional to the non-holomorphic modification
(\ref{eq:free-energy-Omega}) of the free energy. However,
$\hat{\mathcal{H}}$ is by definition a function of the shifted
$y_I$-variables $\hat{y}_I = y_I + i (\Omega_I - \Omega_{\bar I})$.
When using $(x^I, \hat{y}_I, \Upsilon, {\bar \Upsilon})$ as the
independent variables, the variational principle and the attractor
equations take the same form as before.

This observation fits with what is known about the complex
and real polarization for the topological string. The holomorphic
anomaly, which implies the existence of non-holomorphic
modifications of the genus-$g$ topological free energies, is
related to the fact that a complex parametrization of the
moduli space requires the explicit choice of a complex
structure \cite{Witten:1993ed}.
If one instead chooses to parametrize the moduli space by
real period vectors (the real polarization), then no explicit
choice of a complex structure is required, and one arrives
at a `background independent' formulation \cite{Verlinde:2004ck}.
Note, however, that the non-holomorphic terms of the complex
parametrization are encoded in certain non-harmonic terms
in the real parametrization.

Observe that there exists a two-form $\omega= \de x^I \wedge \de
y_I$, which in special geometry is the symplectic form associated with
the flat Darboux coordinates $(x^I, y_I)$. This form is invariant under
electric/magnetic duality. Possible $R^2$-corrections leave this
two-form unaltered, whereas, in the presence of non-holomorphic
corrections one expects that the appropriate extension will be given
by $\omega= \de x^I \wedge \de \hat y_I$. The implication of
this extension is not fully known, but this observation will play a
role later on.

\section{Partial Legendre transforms and duality}
\label{sec:partial-L-duality}
\setcounter{equation}{0}
It is, of course, possible to define the macroscopic entropy as a
Legendre transform with respect to only a subset of the fields, by
substituting a subset of the attractor equations. This subset must be
chosen such that the variational principle remains valid.  These
partial Legendre transforms constitute a hierarchy of Legendre
transforms for the black hole entropy. We discuss two relevant
examples, namely the one proposed in \cite{Ooguri:2004zv}, where all
the magnetic attractor equations are imposed, and the dilatonic one
for heterotic black holes, where only two real potentials are left
which together define the complex dilaton field
\cite{LopesCardoso:1999ur}. At this stage, there is clearly no reason
to prefer one version over the other. This will change in
section~\ref{sec:partition-functions} where we discuss corresponding
partition functions and inverse Laplace transforms for the microscopic
degeneracies.

One possible disadvantage of considering partial Legendre transforms
is that certain invariances are no longer manifest. As it turns
out, the dilatonic formulation does not suffer from this. The
invariances of the dilatonic formulation are relegated to an
additional subsection \ref{sec:T-S-duality}, where we also collect
some useful formulae that we need in later sections.

Let us start and first impose the magnetic attractor equations so that
only the real parts of the $Y^I$ will be relevant. Hence one makes the
substitution,
\begin{equation}
  \label{eq:yphi}
Y^I = \tfrac1{2}(\phi^I + \mathrm{i} p^I) \;.
\end{equation}
The entropy function (\ref{eq:Sigma-simple}) then takes the
form (for the moment we suppress non-holomorphic corrections),
\begin{equation}
  \label{eq:osv1}
  \Sigma(\phi,p,q) =  \mathcal{F}_{\rm
  E}(p,\phi,\Upsilon,\bar\Upsilon) - q_I   \,\phi^I  \,,
\end{equation}
where the corresponding free energy $\mathcal{F}_{\rm E}(p,\phi)$ equals
\begin{equation}
  \label{eq:osv2}
{\cal F}_{\rm E}(p,\phi,\Upsilon,\bar\Upsilon)  = 4  \,{\rm Im} \Big[F
(Y,\Upsilon)\Big]_{Y^I=(\phi^I+\im p^I)/2}  \;.
\end{equation}
To show this one makes use of the homogeneity of the function
$F(Y,\Upsilon)$.

When extremizing (\ref{eq:osv1}) with respect to $\phi^I$ we obtain
the attractor equations $q_I= \partial \mathcal{F}_{\rm E}/\partial
\phi^I$. This shows that the macroscopic entropy is a Legendre
transform of $\mathcal{F}_{\rm E}(p,\phi)$ subject to $\Upsilon=-64$,
as was first noted in \cite{Ooguri:2004zv}. The existence of this
transformation motivated the conjecture that there is a relation with
topological strings, in view of the fact that $\exp[\mathcal{F}_{\rm
  E}]$ equals the modulus square of the topological string partition
function (c.f. (\ref{eq:toplogical})).

Let us now introduce the non-holomorphic corrections to the above
result, by starting from the entropy function (\ref{eq:Sigma-Omega})
and comparing to (\ref{eq:osv1}). This leads to a modification of the
expression (\ref{eq:osv2}) for $\mathcal{F}_{\rm E}(p,\phi)$
\cite{Cardoso:2004xf},\footnote{
  Observe that this result cannot just be obtained by replacing the
  holomorphic function $F(Y,\Upsilon)$ by $F(Y,\Upsilon)+ 2\im
  \,\Omega$. }  
\begin{equation}
  \label{eq:osv2-nonholo}
  \mathcal{F}_{\rm E}(p,\phi) = 4\,\Big[ {\rm Im}\,F(Y,\Upsilon) +
  \Omega(Y,\bar Y,\Upsilon,\bar\Upsilon)\Big]_{Y^I=(\phi^I+ \im
  p^I)/2}  \,.
\end{equation}
The form of the attractor equations, $q_I= \partial \mathcal{F}_{\rm
  E}/\partial \phi^I$, remains unchanged and is equivalent to the second
equation of (\ref{eq:nonholoattr}). Note, however, that the electric
and magnetic charges have been treated very differently in this case,
so that duality invariances that involve both types of charges are
hard to discuss.

Along the same line one can now proceed and eliminate some of the
$\phi^I$ as well. A specific example of this, which is relevant in
later sections, is the dilatonic formulation for heterotic black
holes, where we eliminate all the $\phi^I$ with the exception of two
of them which parametrize the complex dilaton field. This leads to an
entropy function that depends only on the charges and on the dilaton
field \cite{LopesCardoso:1999ur,Cardoso:2004xf}. We demonstrate some salient
features below and in the next subsection~\ref{sec:T-S-duality}. Here
it is convenient to include all the $\Upsilon$-dependent terms into
$\Omega$, which also contains the non-holomorphic corrections. The
heterotic classical function $F(Y)$ is given by
\begin{equation}
  \label{eq:het-F0}
  F(Y) = - \frac{Y^1\,Y^a\eta_{ab} Y^b}{Y^0}\;,\qquad a = 2,\ldots,n,
\end{equation}
with real constants $\eta_{ab}$. The function $\Omega$ depends only
linearly on $\Upsilon$ and $\bar\Upsilon$, as well as on the complex
dilaton field $S= -\im\,Y^1/Y^0$ and its complex conjugate $\bar S$.
Imposing all the magnetic attractor equations yields,
\begin{equation}
  \label{eq:hetcalf}
  {\cal F}_{\rm E}(p,\phi) = \ft12(S + {\bar S}) \left(
    p^a \eta_{ab} p^b - \phi^a\eta_{ab} \phi^b \right)
    - \im (S - {\bar S}) \, p^a \eta_{ab} \phi^b  + 4\,\Omega(S,\bar
      S,\Upsilon,\bar\Upsilon) \,,
\end{equation}
where the dilaton field is expressed in the remaining fields $\phi^0$
and $\phi^1$ and the charges $p^0, p^1$ according to
\begin{equation}
  \label{eq:dilaton}
  S= \frac{- \im \phi^1 + p^1} {\phi^0+ \im p^0}\;.
\end{equation}

Subsequently we impose the electric attractor equations for the $q_a$,
which leads to a full determination of the $Y^I$ in terms of the
dilaton field,
\begin{equation}
  \label{eq:Ya-attractors}
  Y^0= \frac{\bar P(\bar S)}{S+\bar S} \;,\qquad
  Y^1 = \frac{\im S\, \bar P(\bar S)}{S+\bar S} \;,\qquad
 Y^a= - \frac{\eta^{ab}\bar Q_b(\bar S)}{2(S+\bar S)} \;, 
\end{equation}
where we used $\eta^{ac}\, \eta_{cb} = \delta^a{}_b$, and we
introduced the quantities, 
\begin{eqnarray}
  \label{eq:P-Q}
  Q(S) &=& q_0+\mathrm{i}Sq_1 \,,\nonumber\\ 
  P(S)&=& p^1-\mathrm{i}Sp^0\,,\nonumber\\
  Q_a(S)&=& q_a+2\mathrm{i}S\,\eta_{ab} p^b \,.
\end{eqnarray}
Now the entropy function equals
\begin{equation}
  \label{eq:Sigma-dil}
  \Sigma(S,\bar S, p^I, q_I) = \mathcal{F}_{\rm D}(S,\bar S,p^I,q_a) -
  q_0\phi^0 -q_1\phi^1\,,
\end{equation}
where
\begin{eqnarray}
  \label{eq:F-D}
   \mathcal{F}_{\rm D}(S,\bar S,p^I,q_a) &=&  \mathcal{F}_{\rm
   E}(p,\phi)  - q_a\phi^a\nonumber\\
   &=& {}
   \frac{\tfrac1{2} q_a\eta^{ab}q_b +\im\,p^aq_a (S-\bar S) +
   2\,\vert S\vert^2 \,p^a\eta_{ab}p^b}{S+\bar S}  + 4\,\Omega(S,\bar
   S,\Upsilon,\bar\Upsilon) \,,
\end{eqnarray}
and
\begin{equation}
  \label{eq:q-phi}
   q_0\phi^0 +q_1\phi^1= \frac{ 2\,q_0p^1
   -\im(q_0p^0-q_1p^1) (S-\bar S) - 2\,q_1p^0\,\vert S\vert^2 }{S+\bar
   S} \;.
\end{equation}
Combining (\ref{eq:F-D}) with (\ref{eq:q-phi}) yields,
\begin{eqnarray}
\label{eq:nonholoSigma}
  \Sigma(S,\bar S,p,q) =
  - \frac{q^2 - \im p\cdot q \, (S - {\bar S}) + p^2\,|S|^2}
    {S + {\bar S}}  + 4\, \Omega(S,\bar S,\Upsilon,\bar\Upsilon)\;,
\end{eqnarray}
where $q^2$, $p^2$ and $p\cdot q$ are T-duality invariant bilinears of
the various charges, defined by
\begin{eqnarray}
   \label{eq:chargeinvariants}
q^2 &=&  2 q_0 p^1 - \ft12  q_a \eta^{ab} q_b  \;,\nonumber\\
p^2 &=& - 2  p^0 q_1 - 2 p^a \eta_{ab} p^b
 \;, \nonumber\\
q\cdot p &=& q_0 p^0 - q_1 p^1 +  q_a p^a \;.
\end{eqnarray}
These are the expressions that were derived in
\cite{LopesCardoso:1999ur,Cardoso:2004xf}. The remaining attractor equations
coincide with $\partial_S\Sigma(S,\bar S,p,q)=0$,
\begin{equation}
  \label{eq:attractor-S}
  \frac{q^2 + 2\im \,p\cdot q \, {\bar S} - p^2\,\bar S^2}{S+\bar S}
  + 4\,(S+\bar S)\partial_S\Omega = 0\,.
\end{equation}
Provided that $\Omega$ is invariant under S-duality, all the above
equations are consistent with S- and T-duality as can be verified by
using the transformation rules presented in the subsection below. As
before, the value of $\Sigma(S,\bar S,p,q)$ at the attractor point
(including $\Upsilon=-64$) will yield the macroscopic entropy as a
function of the charges. Also the entropy will then be invariant under
T- and S-duality.

Finally we consider the quantity $\hat K$,
\begin{eqnarray}
  \label{eq:K-hat}
  \hat K &=& \im (\bar Y^I\hat F_I - Y^I \bar{\hat F}_I ) \nonumber\\
  &=&{}
  \vert Y^0\vert^2 (S+\bar S)\left[ (T+\bar T)^a\eta_{ab} (T+\bar T)^b
  + \frac{2\,\partial_S\Omega}{(Y^0)^2}
  + \frac{2\,\partial_{\bar S}\Omega}{(\bar Y^0)^2} \right] \,,
\end{eqnarray}
where the S-duality invariant moduli $T^a$ are defined by $T^a= -\im
\,Y^a/Y^0$. Note that, by construction, $\hat K$ is invariant under T-
and S-duality, as can be verified by using the transformations given
in the subsection below.\footnote{
Note that the invariance under T-duality is somewhat
more subtle, as one can deduce immediately
from the transformation of $T^a$ under T-duality,
\begin{equation}
  \label{eq:T-dual-T}
  \delta T^a= \im\,b^a + c\,T^a + \im\,a_b\Big[ -\ft12
  \eta^{ab}\left(T^c\eta_{cd}T^d +
  2\,(Y^0)^{-2}\partial_S\Omega\right)  + T^a\,T^b   \Big]\,. \nonumber
\end{equation}
}  At the attractor point $\hat K$ is
proportional to the area, as follows from,
\begin{equation}
  \label{eq:area-heterotic}
  \hat K\Big\vert_{\rm attractor} = \vert Z\vert^2 =   - \frac{q^2 - \im
  p\cdot q \, (S - {\bar S}) + p^2\,|S|^2} {S + {\bar S}}  \,,
\end{equation}
subject to the attractor equation (\ref{eq:attractor-S}).

Heterotic BPS black holes can either be large or small. Small black
holes have vanishing area at the two-derivative level, and they
correspond to electrically charged 1/2-BPS states. When taking
$R^2$-interactions into account, a horizon forms and also the entropy
becomes non-vanishing. This phenomenon has been studied in more detail
in \cite{Dabholkar:2004dq,Sen:2004dp,Hubeny:2004ji}. Large black
holes, on the other hand, have non-vanishing area at the
two-derivative level.  In models with $N=4$ supersymmetry, they
correspond to dyonic 1/4-BPS states.
\subsection{Duality invariance and non-holomorphic corrections}
\label{sec:T-S-duality}
In this subsection we demonstrate how the non-holomorphic terms enter
in order to realize the invariance under certain duality symmetries.
Here we follow the same strategy as in \cite{LopesCardoso:1999ur}, but
will consider a more extended class of models with $N=4$
supersymmetry. In this work the language of $N=2$ supergravity was
used to establish the invariance under target-space duality
(T-duality) and S-duality of black holes with $N=4$ supersymmetry that
arise in the toroidal compactification of heterotic string theory.
This compactification leads to an effective $N=4$ supergravity coupled
to 22 abelian vector supermultiplets.  Together with the 6 abelian
graviphotons this leads to a total of 28 vector fields. As 4 of the
graviphotons are absent in the truncation to $N=2$ supergravity, the
variables $Y^I$ will be labeled by $I=0,1,\ldots,23$. The central idea
is to determine the entropy function in the context of $N=2$
supergravity and to extend the charges at the end to 28 electric and
28 magnetic charges, by making use of T- and S-duality. However, there
exist other $N=4$ heterotic models based on modding out the theory by
the action of some discrete abelian group, which can be discussed on a
par. They correspond to a class of so-called CHL models
\cite{Chaudhuri:1995fk}, which have fewer than 28 abelian gauge
fields.  The $N=2$ description is then based on a smaller number of
fields $Y^I$, which we will specify in due course. At symmetry
enhancement points in the respective moduli spaces the abelian gauge group
is enlarged to a non-abelian one. All these models are dual to certain
type-IIA string compactifications.

In the following we will discuss T- and S-duality for this class of
models and describe their entropy functions.  The $N=2$ description is
based on the holomorphic function (\ref{eq:het-F0}), modified with
$\Upsilon$-dependent terms and possibly non-holomorphic corrections
encoded in the function $\Omega$.  In this case $\Omega$ depends only
on $\Upsilon$ and on the dilaton field $S$, and their complex
conjugates,
\begin{equation}
  \label{eq:het-F}
  F  = - \frac{Y^1\,Y^a\eta_{ab}Y^b}{Y^0} + 2\im
  \,\Omega(S,\bar S,\Upsilon,\bar\Upsilon)
\;, \qquad a = 2, \dots, n \,,
\end{equation}
where the dilaton-axion field is described by $S =- \mathrm{i} \,
Y^1/Y^0$, and $\eta_{ab}$ is an $\mathrm{SO}(1,n-2)$ invariant metric
of indefinite signature. The number $n$ denotes the number of moduli
fields, and is left unspecified for the time being. It is related to
the rank of the gauge group that arises in the $N=4$ compactification.
The $\hat F_I$ associated with (\ref{eq:het-F}) are given by
\begin{eqnarray}
  \label{eq:F_I}
  \hat F_0&=&  \frac{Y^1}{(Y^0)^2} \Big[Y^a\eta_{ab}Y^b
   -2\,\partial_S\Omega \Big]   \,,\nonumber\\
  \hat F_1&=& - \frac{1}{Y^0} \Big[ Y^a\eta_{ab}Y^b -2\,\partial_S\Omega
  \Big]  \,, \nonumber\\
  \hat F_a&=& - \frac{2\,Y^1}{Y^0} \, \eta_{ab}Y^b \,,
\end{eqnarray}
where we note the obvious constraint $Y^0\hat F_0+Y^1\hat F_1=0$.

We now investigate under which condition the above function leads to
T- and S-duality. In the case of a holomorphic function the period
vector $(Y^I, F_I)$ transforms in the usual way under symplectic
transformations induced by electric/magnetic duality.  When a subgroup
of these symplectic transformations constitutes an invariance of
the Wilsonian action, this implies that the transformations of the
$Y^I$ will induce precisely the correct transformations on the $F_I$. In
the case of non-holomorphic terms one would like this to remain true
so that the attractor equations will be consistent with the duality
invariance. Following \cite{LopesCardoso:1999ur} we first turn to
T-duality, whose infinitesimal transformations are given by
\begin{equation}
    \label{eq:tduality}
    \begin{array}{rcl}
\delta Y^0 &=& - c \, Y^0 - a_a \, Y^a \;,\\
\delta Y^1 &=& - c \, Y^1 + \ft12 \eta^{ab}
a_a \, \hat F_b  \;,\\
\delta Y^a &=& - b^a \, Y^0 + \ft12 \eta^{ab} a_b \,\hat F_1  \;,
    \end{array}
    \begin{array}{rcl}
  \delta \hat F_0 &=& c\, \hat F_0 + b^a \, \hat F_a \;, \\
  \delta \hat F_1 &=& c \, \hat F_1 + 2 \eta_{ab} b^ a \, Y^b \;,\\
  \delta \hat F_a &=& a_a \, \hat F_0 + 2 \eta_{ab} b^b \, Y^1 \;,
    \end{array}
\end{equation}
where the $a_a$, $b^a$ and $c$ denote $2n-1$ infinitesimal
transformation parameters; upon combining these transformations with
the obvious $\mathrm{SO}(1,n-2)$ transformations that act linearly on
the $Y^a$ (and on the $\hat F_a$), one obtains the group
$\mathrm{SO}(2,n-1)$. Note that the dilaton field $S$ is invariant
under T-duality, while $(Y^0,\, \hat F_1,\, Y^a)$ and $(\hat
F_0,-Y^1,\, \hat F_a)$ transform both as vectors under
$\mathrm{SO}(2,n-1)$. It can now be verified that the variations
$\delta \hat F_I$ are precisely induced by the variations $\delta
Y^I$, irrespective of the precise form of $\Omega(S,\bar
S,\Upsilon,\bar\Upsilon)$.

Under {\it finite}  S-duality transformations, the situation is more
complicated. Here the $Y^I$ transform as follows,
\begin{eqnarray}
  \label{eq:S-Y}
  Y^0 &\to&  d\,Y^0 + c\, Y^1 \,,\nonumber\\
  Y^1 &\to&  a\,Y^1 + b\, Y^0 \,, \nonumber\\
  Y^a &\to&  d\,Y^a -\ft12 c\,\eta^{ab}\,\hat F_b \,,
\end{eqnarray}
where $a,b,c,d$ are integers, or belong to a subset of integers that
parametrize a subgroup of $\mathrm{SL}(2;\mathbb{Z})$, and satisfy
$ad-bc=1$. As a result of these transformations, $S$ transforms
according to the well-known formulae,
\begin{equation}
  \label{eq:Strafo}
  S \rightarrow
  S^\prime = \frac{ a S - \im b}{\im c S + d} \;, \qquad \frac{\partial
  S^\prime}{\partial S} = \frac{1}{( \mathrm{i} c \, S + d )^2} \;.
\end{equation}
When applied to the $\hat F_I$ these transformations induce the changes,
\begin{eqnarray}
  \label{eq:S-F}
  \hat F_0 &\to& a\, \hat F_0 - b\,\hat F_1 + \Delta_0 \,,\nonumber \\
  \hat F_1 &\to& d\,\hat F_1 - c\,\hat F_0 + \Delta_1 \,, \nonumber \\
  \hat F_a &\to& a\,\hat F_a -2b \,\eta_{ab} Y^b\,,
\end{eqnarray}
where $\Delta_0$ and $\Delta_1$ are proportional to the same
expression,
\begin{eqnarray}
  \label{eq:S-Delta}
  \Delta_0 \propto\Delta_1 \propto
  \partial_{S^\prime} \Omega(S^\prime,{\bar S}^\prime,
  \Upsilon,\bar\Upsilon) -    (\im c \,S+d )^2 \,\partial_S
  \Omega(S,{\bar S},\Upsilon,\bar\Upsilon)\;,
\end{eqnarray}
which vanishes when $\partial_S\Omega$ is a modular form of weight two
\cite{LopesCardoso:1999ur}. However, it is well known that there
exists no modular form of weight two. In order to have attractor
equations that transform covariantly under S-duality we are therefore
forced to include non-holomorphic expressions. In applying this
argument one may have to restrict $\Upsilon$ to its attractor value,
but subject to this restriction $\Omega$ must be invariant under
S-duality. Once the S-duality group is specified, the form of $\Omega$
will usually follow uniquely. 

Observe that the duality transformations of the charges follow
directly from those of the periods. In particular, the charge vectors
$(p^0,q_1,p^a)$ and $(q_0,-p^1,q_a)$ transform irreducibly under the
T-duality group. The three T-duality invariants (\ref{eq:chargeinvariants})
transform as a vector under the S-duality group. Furthermore the
quantities (\ref{eq:P-Q}) transform under S-duality as a modular
function, 
\begin{equation}
  \label{eq:S-PQQ}
  (Q(S),\, P(S),\, Q_a(S)) \longrightarrow  \frac{1}{\im c\,S+d} \,
  (Q(S), \,P(S),\, Q_a(S)) \,,
\end{equation}
and as a vector under T-duality. 

We will now discuss the expressions for $\Omega$ for the class of CHL
models \cite{Chaudhuri:1995fk} discussed recently in
\cite{Sen:2005pu,Sen:2005ch,Jatkar:2005bh}. First we introduce the unique
cusp forms of weight $k+2$ associated with the S-duality group
$\Gamma_1(N)\subset\mathrm{SL}(2;\mathbb{Z})$, defined by
\begin{equation}
  \label{eq:Gamma-cusp}
  f^{(k)}(S) = \eta^{k+2}(S)\;\eta^{k+2}(N S)\,,
\end{equation}
where $N$ is a certain positive integer. Hence these cusp forms
transform under the S-duality transformations that belong to the
subgroup $\Gamma_1(N)$, according to
\begin{equation}
  \label{eq:f-S-dual}
  f^{(k)} (S^\prime) = (\im c \,S+d )^{k+2}  \, f^{(k)}(S)\,.
\end{equation}
The subgroup $\Gamma_1(N)$ requires  the transformation parameters to be
restricted according to $c = 0\mod N$ and $a,d=1\mod N$, which is
crucial for deriving the above result. The integers $k$ and $N$
are not independent in these models and subject to
\begin{equation}
  \label{eq:k-N}
  (k+2) (N+1) =24 \,.
\end{equation}
The values $k=10$ and $N=1$ correspond to the toroidal
compactification. Following \cite{Jatkar:2005bh}, we will restrict
attention to the values $(N,k)= (1,10), (2,6), (3,4), (5,2)$ and
$(7,1)$. The rank of the gauge group (corresponding to the number of
abelian gauge fields in the effective supergravity action) is then
equal to $r= 28, 20, 16, 12$ or $10$, respectively. The corresponding
number of $N=2$ matter vector supermultiplets is then given by 
$n=2(k+2)-1$. 

The function $\Omega$ can now be expressed in terms of the cusp forms,
\begin{equation}
  \label{eq:Omega-het}
  \Omega_k(S,\bar S,\Upsilon,\bar\Upsilon) =
  {}\frac{1}{256\,\pi} \Big[\Upsilon \log f^{(k)}(S)  +
\bar\Upsilon \log f^{(k)}(\bar S) + \ft12(\Upsilon+\bar \Upsilon)
\log (S+\bar S)^{k+2} \Big] \,,
\end{equation}
in close analogy to the case $k=10$
\cite{LopesCardoso:1999ur,Cardoso:2004xf}. Note that these terms agree
with the terms obtained for the corresponding effective actions (see,
for instance, \cite{Harvey:1996ir,Gregori:1996}). Suppressing
instanton corrections this result takes the form,
\begin{equation}
  \label{eq:Omega-asymp}
  \Omega_k(S,\bar S,\Upsilon,\bar\Upsilon)\Big\vert_{\Upsilon =
  - 64}
  = \ft12 (S+\bar S) - \frac{k+2}{4\pi}  \log(S+\bar S)\,.
\end{equation}
This implies that, in the limit of large charges, the entropy of small
black holes (with vanishing charges $p^0,q_1,p^2,\ldots, p^n$) will be
independent of $k$ and its leading contribution will be equal to
one-half of the area. The latter follows from the entropy function
(\ref{eq:nonholoSigma}) which, in this case, reads,
\begin{equation}
  \label{eq:electric-Sigma}
  \Sigma(S,\bar S,p,q) = - \frac{q^2}{S+\bar S} + 2 (S+\bar S) -
  \frac{k+2}{\pi} \log(S+\bar S)\,.
\end{equation}
Stationarity of $\Sigma$ shows that $S+\bar S\approx \ft1{4\pi}(k+2) +
\sqrt{\vert q^2\vert/2}$, while the entropy $S_{\mathrm{macro}}\approx
4\pi\sqrt{\vert q^2\vert/2} -\ft12(k+2)\log\vert q^2\vert$. The
logarithmic term is related to the non-holomorphic term in
(\ref{eq:Omega-het}), and its 
coefficient is not in agreement with microstate counting. However, this
term is subject to semiclassical corrections, as we will discuss in
the following sections. In the same approximation the area equals
$8\pi\sqrt{\vert q^2\vert/2}$. 

\section{Partition functions and inverse Laplace transforms}
\label{sec:partition-functions}
\setcounter{equation}{0}
So far, we discussed black hole entropy from a macroscopic point of
view. To make the connection with microstate degeneracies, we
conjecture, in the spirit of \cite{Ooguri:2004zv}, that the Legendre
transforms of the entropy are indicative of a thermodynamic origin of
the various entropy functions. It is then natural to assume that the
corresponding free energies are related to black hole partition
functions corresponding to suitable ensembles of black hole
microstates. To examine the consequences of this idea, let us define
the following partition function,
\begin{equation}
  \label{eq:partition}
  Z(\phi,\chi) = \sum_{\{p,q\}} \;   d(p,q)
  \, \mathrm{e}^{\pi [q_I \phi^I - p^I \chi_I]} \,,
\end{equation}
where $d(p,q)$ denotes the microscopic degeneracies of the black hole
microstates with black hole charges $p^I$ and $q_I$. This is the
partition sum over a canonical ensemble, which is invariant under the
various duality symmetries, provided that the electro- and
magnetostatic potentials $(\phi^I,\chi_I)$ transform as a symplectic
vector. Identifying a free energy with the logarithm of $Z(\phi,\chi)$
it is clear that it should, perhaps in an appropriate limit, be
related to the macroscopic free energy introduced earlier. On the
other hand, viewing $Z(\phi,\chi)$ as an analytic function in $\phi^I$
and $\chi_I$, the degeneracies $d(p,q)$ can be retrieved by an inverse
Laplace transform,
\begin{equation}
  \label{eq:inverse-LT}
  d(p,q) \propto \int \;{\de\chi_I\,\de\phi^I}\;
  Z(\phi,\chi) \; \ee^{\pi [-q_I \phi^I + p^I \chi_I]} \,,
\end{equation}
where the integration contours run, for instance, over the intervals
$(\phi-\im, \phi +\im)$ and $(\chi-\im, \chi +\im)$ (we are assuming
an integer-valued charge lattice). Obviously, this makes sense as
$Z(\phi,\chi)$ is formally periodic under shifts of $\phi$ and $\chi$
by multiples of $2\im$.

All of the above arguments suggest to identify $Z(\phi,\chi)$ with
the generalized Hesse potential, introduced in
subsection~\ref{sec:hesse},  
\begin{equation}
  \label{eq:partition-hesse}
  \sum_{\{p,q\}} \;   d(p,q)
  \,\mathrm{e}^{\pi[q_I \phi^I - p^I \hat \chi_I]} \sim 
  \sum_{\rm shifts} \;
  \mathrm{e}^{2\pi\,\mathcal{H}(\phi/2,\hat\chi/2,\Upsilon,\bar\Upsilon)}
  \,,
\end{equation}
where $\Upsilon$ is equal to its attractor value and where the
definition of $\hat \chi= 2\,\hat y$ was explained in subsection
\ref{sec:hesse}.  Because the generalized Hesse potential is a
macroscopic quantity which does not in general exhibit the periodicity
that is characteristic for the partition function, the right-hand side
of (\ref{eq:partition-hesse}) requires an explicit periodicity sum
over discrete imaginary shifts of the $\phi$ and $\chi$. In case that
the Hesse potential exhibits a certain periodicity (say, with a
different periodicity interval), then the sum over the imaginary
shifts will have to be modded out appropriately such as to avoid
overcounting. This is confirmed by the result of the calculation we will
present in subsection~\ref{sec:sec:next-dyonic}. 
At any rate, we expect
that when substituting $2\pi\mathcal{H}$ into the inverse Laplace
transform, the periodicity sum can be incorporated into the integration
contours.

Unfortunately, it is in general difficult to find an explicit
representation for the Hesse potential, as the relation (\ref{eq:u-y})
between the complex variables $Y^I$ and the real variables $x^I$ and
$y_I$ is complicated. Therefore we rewrite the above formulae in terms
of the complex variables $Y^I$, where explicit results are known. In
that case the relation (\ref{eq:partition-hesse}) takes the following
form,
\begin{equation}
  \label{eq:partition1}
  \sum_{\{p,q\}} \;   d(p,q)
  \, \mathrm{e}^{\pi [q_I (Y+\bar Y)^I - p^I (\hat F +\hat{\bar
  F})_I]} \sim 
  \sum_{\rm shifts} \;
  \ee^{\pi \,\mathcal{F} (Y,\bar Y, \Upsilon,\bar \Upsilon) } \,,
\end{equation}
where $\mathcal{F}$ equals the free energy
(\ref{eq:free-energy-Omega}). Here we note that according to
(\ref{eq:Sigma-variation-1}) the natural variables on which
$\mathcal{F}$ depends, are indeed the real parts of $Y^I$ and $\hat
F_I$.  Needless to say, the relation (\ref{eq:partition1}) (and its
preceding one) is rather subtle, but it is reassuring that both sides
are manifestly consistent with duality.

Just as indicated in (\ref{eq:inverse-LT}), it is possible to formally
invert (\ref{eq:partition1}) by means of an inverse Laplace transform,
\begin{eqnarray}
  \label{eq:laplace1}
  d(p,q) &\propto& \int \; \de (Y+\bar Y)^I\; \de (\hat F+\hat{\bar F})_I
  \;\ee^{\pi\,\Sigma(Y,\bar Y,p,q)} \nonumber \\
  &\propto& \int \; \de Y\, \de\bar Y\;\Delta^-(Y,\bar Y)\;
  \ee^{\pi\,\Sigma(Y,\bar Y,p,q)}\;,
\end{eqnarray}
where $\Delta^-(Y,\bar Y)$ is an integration measure whose form
depends on $\hat F_I+ {\hat {\bar F}}_I$. The expression for
$\Delta^-$ follows directly from (\ref{eq:hat-F-I}). We also define a
similar determinant $\Delta^+$ that we shall need shortly,
\begin{equation}
  \label{eq:measure}
  \Delta^\pm(Y,\bar Y) = \left\vert\det\Big[{\rm Im}\, F_{KL} + 2
  \,{\rm Re}(\Omega_{KL} \pm \Omega_{K{\bar L}}) \Big]\right\vert \,. 
\end{equation}
Here we note the explicit dependence on $\Omega$. As before, $F_{IJ}$
and $F_I$ refer to $Y$-derivatives of the holomorphic function
$F(Y,\Upsilon)$ whereas $\Omega_{IJ}$ and $\Omega_{I\bar J}$ denote the
holomorphic and mixed holomorphic-antiholomorphic second derivatives
of $\Omega$, respectively. 

It is not a priori clear whether the integral (\ref{eq:laplace1}) is
well-defined. Although the integration contours can in principle be
deduced from the contours used in (\ref{eq:inverse-LT}), an explicit
determination is again not possible in general. Of course, the
contours can be deformed but this depends crucially on the integrand
whose analytic structure is a priori not clear. Here, it is important to
realize that the analytic continuation refers to the initial variables
in the inverse Laplace transform (\ref{eq:inverse-LT}), provided by
the electro- and magnetostatic potentials. Therefore the analytic
continuation does not automatically respect the relation between $Y^I$
and $\bar Y^I$ based on complex conjugation. 
Just as before, the effect of
a periodicity sum on the right-hand side of (\ref{eq:partition1}) can
be incorporated into the integration contour, but the periodicity sum
is also defined in terms of the original variables. Obviously these
matters are rather subtle and can only be addressed in specific
models. A separate requirement is that the integration contours should
be consistent with duality. Here it is worth pointing out that
explicit integral representations for microscopic black hole
degeneracies are known, although their (auxiliary) integration
parameters have no direct macroscopic significance, unlike in
(\ref{eq:laplace1}). These representations will shortly play an
important role.

Leaving aside these subtle points we will first establish that the
integral representation (\ref{eq:laplace1}) makes sense in case that a
saddle-point approximation is appropriate. In view of the previous
results it is clear that the saddle point coincides with the attractor
point, so that the integrand should be evaluated on the attractor point.
Evaluating the second variation of $\Sigma$,
\begin{eqnarray}
  \label{eq:var2-Sigma-Omega}
  \delta^2 \Sigma &=& \mathrm{i} (Y^I - {\bar Y}^I -i p^I)\,
  \delta^2 (F_I+\bar F_I + 2\mathrm{i}(\Omega_I- \Omega_{\bar I}))
  \nonumber \\
&&{}
  + 2\mathrm{i} \left(\delta Y^I \,\delta (\bar F_I - 2\mathrm{i}
  \Omega_{\bar I})
  - \delta(F_I +2\mathrm{i} \Omega_I)\,\delta \bar Y^I \right) \,,
\end{eqnarray}
and imposing the attractor equations so that $\delta\Sigma=0$, one
expands the exponent around the saddle point and evaluates the
semiclassical Gaussian integral. This integral leads to a second
determinant which, when $Y^I-\bar Y^I-\mathrm{i}p^I=0$, factorizes
into the square roots of two subdeterminants, $\sqrt{\Delta^+}$ and
$\sqrt{\Delta^-}$. Here the plus (minus) sign refers to the
contribution of integrating over the real (imaginary) part of $\delta
Y^I$.  Consequently, the result of a saddle-point approximation
applied to (\ref{eq:laplace1}) yields,
\begin{equation}
  \label{eq:complex-saddle}
  d(p,q) = \sqrt{
  \left\vert \,\frac{\Delta^-(Y,\bar Y)}{\Delta^+(Y,\bar Y)}\right\vert }_{\rm
   attractor} \; \ee^{{\cal S}_{\rm macro}(p,q)} \;.
\end{equation}
In the absence of non-holomorphic corrections the ratio of the two
determinants is equal to unity and one thus recovers precisely the
macroscopic entropy. Furthermore one can easily convince oneself that
the saddle-point approximation leads to results that are compatible
with duality. 

Before discussing this result in more detail, let us also consider the
case where one integrates only over the imaginary values of $\delta
Y^I$ in saddle-point approximation. The saddle point then occurs in
the subspace defined by the magnetic attractor equations, so that one
obtains a modified version of the OSV integral \cite{Ooguri:2004zv},
\begin{equation}
  \label{eq:laplace-osv}
  d(p,q) \propto \int \; \de \phi \; \sqrt{\Delta^-(p,\phi)} \;
\ee^{\pi[\mathcal{F}_{\rm E}(p,\phi)- q_I\phi^I]} \;,
\end{equation}
where $\mathcal{F}_{\rm E}(p,\phi)$ was defined in
(\ref{eq:osv2-nonholo}) and $\Delta^-(p,\phi)$ is defined in
(\ref{eq:measure}) with the $Y^I$ given by (\ref{eq:yphi}). Hence this
integral must contain a measure factor $\sqrt{\Delta^-}$ in order to
remain consistent with electric/magnetic duality.\footnote{
  There has been some discussion in the literature about a possible
  modification of this integral, such as for instance by a measure
  factor \cite{Dabholkar:2005dt,Shih:2005he,Parvizi:2005aa}. See also
  footnote~\ref{footnote1}. } 
Without the measure factor this is the integral conjectured by
\cite{Ooguri:2004zv}. 
Inverting this formula to a partition sum over a mixed ensemble, one
finds,
\begin{equation}
  \label{eq:partition1-osv} Z(p,\phi)=
    \sum_{\{q\}} \; d(p,q) \, \mathrm{e}^{\pi \, q_I \phi^I } \sim
  \sum_{\rm shifts} \;
  \sqrt{\Delta^-(p,\phi)}\;
    \mathrm{e}^{\pi\,\mathcal{F}_{\rm E}(p,\phi)}  \;.
\end{equation}
However, we note that this expression and the preceding one is less
general than (\ref{eq:laplace1}) because it involves a saddle-point
approximation. Moreover the function $\mathcal{F}_{\mathrm{E}}$ is not
duality invariant and the invariance is only recaptured when
completing the saddle-point approximation with respect to the fields
$\phi^I$. Therefore one expects that an evaluation of
(\ref{eq:laplace-osv}) beyond the saddle-point approximation will
entail a violation of (some of) the duality symmetries again, because
it amounts to an unequal treatment of the real and the imaginary parts
of the $Y^I$. Hence the situation regarding (\ref{eq:laplace-osv}) and
(\ref{eq:partition1-osv}) remains unsatisfactory.

In \cite{Ooguri:2004zv}, the partition function $Z(p,\phi)$ was
conjectured to be given by the modulus square of the partition
function of the topological string. This equality holds provided
$\mathcal{F}_{\mathrm{E}}$ does not contain contributions from
non-holomorphic terms, and provided there is no nontrivial
multiplicative factor. The above observation has inspired further
interest in the relation between the holomorphic anomaly equation of
topological string theory and possible contributions from
non-holomorphic terms to the black hole entropy (see the work quoted
in the introduction). However, as we already mentioned in the
introduction, a known relationship exists via the non-Wilsonian part
of the effective action. For instance, as shown in
\cite{Antoniadis:1994}, the holomorphic anomaly of topological string
partition functions is precisely related to the non-local part of the
action induced by massless string states.

Obviously, one can test the underlying conjecture by calculating the
inverse Laplace transforms (\ref{eq:laplace1}) and
(\ref{eq:laplace-osv}), using the macroscopic data as input and
comparing with the known asymptotic degeneracies. Another approach is
to start instead from known microscopic degeneracies and determine the
partition functions (\ref{eq:partition1}) or
(\ref{eq:partition1-osv}), which can then be compared to the
macroscopic data. Unfortunately there are not many examples available
where one knows both macroscopic and microscopic results.  In the
remainder of this paper we will therefore restrict ourselves to the
case of heterotic black holes with $N=4$ supersymmetry, which we
already introduced from a macroscopic perspective in
section~\ref{sec:partial-L-duality}. Although there are positive
results, many intriguing questions remain. A particular pertinent
question concerns the domain of validity of this approach, which,
unfortunately, we will not be able to answer.

In section~\ref{sec:heterotic-bh} we will approach the comparison from
the microscopic side, while in this section we will start from the
macroscopic side and examine a number of results based on the inverse
Laplace transforms (\ref{eq:laplace1}) and (\ref{eq:laplace-osv}). A
number of tests based on (\ref{eq:laplace-osv}), without including the
measure factor $\Delta^-$ and the non-holomorphic contributions, have
already appeared in the literature
\cite{Dabholkar:2004yr,Dabholkar:2005x,Sen:2005ch,Dabholkar:2005dt}.
These concern the electric (small) black holes. However, let us first
discuss the more generic case of large black holes and evaluate the
determinants $\Delta^\pm$ for arbitrary charge configurations.  Some
of the relevant expressions were already presented in
section~\ref{sec:partial-L-duality} and we use them to evaluate the
determinants (\ref{eq:measure}). The result reads as follows,
\begin{equation}
  \label{eq:dets-general}
  \Delta^\pm =  \frac{(S+\bar S)^{n-3}\,\det[-\eta_{ab}] }{4\,\vert
  Y^0\vert^4} \;
    \Big[\left(\hat K \pm 2\,(S+\bar S)^2 \partial_S\partial_{\bar S}
    \Omega \right)^2
    -4\,\left\vert (S+\bar S)^2 \, D_S\partial_S \Omega\right\vert^2 \Big]
  \,,
\end{equation}
where $\hat K$ has been defined in (\ref{eq:K-hat}) and
\begin{equation}
  \label{eq:D-partial-Omega}
  D_S \partial_S \Omega = \partial_S \partial_S \Omega + \frac{2}{S
  + \bar S} \partial_S \Omega \;.
\end{equation}
Provided that $\Omega$ is invariant under S-duality, also $(S+\bar
S)^2 \partial_S\partial_{\bar S} \Omega$ and $\vert (S+\bar S)^2 \,
D_S\partial_S \Omega\vert$ are invariant. This is confirmed by the
fact that the measure $(S+\bar S)^{n-3} \vert Y^0\vert^{-4} \;
\prod_I\, \de Y^I\, \de\bar Y^I$ factorizes into two parts, $[\vert
Y^0\vert^2(S+\bar S)]^{n-1}\,\prod_a\,\de T^a\,\de\bar T^a$, and
$[\vert Y^0\vert(S+\bar S)]^{-2} \,\de Y^0\,\de\bar Y^0 \,\de
S\,\de\bar S$, which are separately S-duality invariant.

In (\ref{eq:area-heterotic}) we established that $\hat K$ equals the
black hole area on the attractor surface. For large black holes one
can take the limit of large charges, keeping the dilaton field finite.
Since $\Omega$ is proportional to $\Upsilon$, it represents subleading
terms. Therefore $\hat K$ yields the leading contribution to the
determinants $\Delta^\pm$, so that the prefactor in the saddle-point
approximation (\ref{eq:complex-saddle}) tends to unity. Hence one
recovers precisely the exponential of the macroscopic entropy. This is
a gratifying result. In the saddle-point approximation the macroscopic
entropy is equal to the logarithm of the microstate degeneracy up to
terms that vanish in the limit of large charges. In the next section
we will consider the opposite perspective and perform a similar
approximation on the formula that encodes the microscopic dyonic
degeneracies which yields exactly the macroscopic result encoded in
(\ref{eq:nonholoSigma}) and (\ref{eq:Omega-het}). Hence the conjecture
leading to (\ref{eq:complex-saddle}) is clearly correct in the
semiclassical approximation.

As it turns out, a similar exercise for electric (small) black holes
leads to a less satisfactory situation, because the generic
saddle-point expression (\ref{eq:complex-saddle}) breaks down. 
This has to do with the vanishing of the determinants.
In general,
the vanishing of the determinant $\Delta^-$ 
implies that the real parts of $(Y^I,\hat F_I)$ are not independent
coordinates and this indicates that the saddle point is not an
isolated point but rather a submanifold of finite dimension at which
the attractor equations will only be partially satisfied.  
At the saddle point $\Delta^\pm$
will vanish whenever the matrix of second derivatives of $\Sigma$ at
the saddle point is degenerate. This is no obstruction to a
saddle-point approximation, but it implies that the general formula
(\ref{eq:complex-saddle}) is no longer applicable.  
For
small black holes the classical contribution to the determinants
$\Delta^\pm$ vanishes at the attractor point, as is clear from
(\ref{eq:attractor-S}) and (\ref{eq:area-heterotic}). So the
subleading corrections are important which tends to make
approximations somewhat unreliable. 
Both these phenomena 
take place when restricting oneself to the classical terms
in the measure and entropy function, and it is therefore clear that the
behaviour of the integral will depend sensitively on the
approximations employed.

Hence we will now perform the saddle-point approximation for `small'
(electric) black holes by following a step-by-step procedure. We
assume that the magnetic attractor equations will be satisfied at the
saddle-point so that we can base ourselves on (\ref{eq:laplace-osv}).
To determine the expressions for $\Delta^\pm$ we first recall that the
charges $p^0,q_1,p^2,\ldots,p^n$ can be set to zero for the electric
case. Therefore the $T$-moduli are equal to $T^a= -\im \phi^a/\phi^0$,
and thus purely imaginary. Consequently they do not contribute to the
expression (\ref{eq:K-hat}) for $\hat K$, and we obtain,
\begin{equation}
  \label{eq:K-hat-electric}
  \hat K_k= 2\,(S+\bar S) (\partial_S\Omega_k + \partial_{\bar
  S}\Omega_k)\,,
\end{equation}
where $k$ labels the particular CHL model.  Substituting this result
into (\ref{eq:dets-general}) yields,
\begin{eqnarray}
  \label{eq:dets-electric}
  \Delta_k^\pm &=& \frac{(S+\bar S)^{n+1}\,\det[-\eta_{ab}]}{(p^1)^4}
  \nonumber\\ 
  &&{} \times
  \Big[\Big( (S+\bar S)(\partial_S+\partial_{\bar S})\Omega_k \pm (S+\bar
  S)^2 \partial_S\partial_{\bar S}\Omega_k\Big)^2 - \left\vert (S+\bar
  S)^2 D_S\partial_S\Omega_k\right\vert^2\Big]\,, 
\end{eqnarray}
which shows that the classical contribution is entirely absent and the
result depends exclusively on $\Omega_k$, defined in
(\ref{eq:Omega-het}). Observe that here and henceforth we take
$\Upsilon=-64$ and suppress the $\Upsilon$-field.

We now turn to the evaluation of the inverse Laplace integral
(\ref{eq:laplace-osv}). First we write down the expression for the
exponent, using (\ref{eq:hetcalf}) and (\ref{eq:dilaton}) and
rewriting $\phi^0$ and $\phi^1$ in terms of $S$ and $\bar S$,
\begin{equation}
  \label{eq:hetcalf-electric}
  {\cal F}_{\rm E}(p,\phi)- q_0 \phi^0- q_a \phi^a = - \ft12(S + {\bar
  S}) 
  \,\phi^a\eta_{ab} \phi^b - q_a\phi^a  - \frac{2q_0p^1}{S+\bar S}  +
  4\,\Omega_k(S,\bar S)  \,.  
\end{equation}
We note that the above expression is not invariant under T-duality.
This is due to the fact that the perturbative electric/magnetic
duality basis counts $p^1$ as an electric and $q_1$ as a magnetic
charge \cite{Ceresole,DKLL}. Following (\ref{eq:laplace-osv}) we
consider the integral,  
\begin{equation}
  \label{eq:osv-electric1}
  d(p^1,q_0,q_a) \propto (p^1)^2 \int
  \frac{\mathrm{d}S\,\mathrm{d}\bar S} {(S+\bar  S)^3} \, \int  
  \prod_{a=2}^n \;\mathrm{d}\phi^a \sqrt {\Delta_k^-(S,\bar S)} \;
  \ee^{\pi[ \mathcal{F}_{\mathrm{E}}- q_0 \phi^0 - q_a\phi^a ]}\, , 
\end{equation}
which is not manifestly T-duality invariant. However, when performing
the Gaussian integrals over $\phi^a$ (ignoring questions of
convergence) we find
\begin{equation}
  \label{eq:osv-electric}
  d(p^1,q_0,q_a) \propto (p^1)^2 \int
  \frac{\mathrm{d}S\,\mathrm{d}\bar S}
  {(S+\bar S)^{(n+5)/2}} 
  \, \sqrt {\Delta_k^-(S,\bar S)} \;
  \exp\Big[ - \frac{\pi\,q^2}{S+\bar S} + 4\pi\,\Omega_k(S,\bar S)\Big] \,,
\end{equation}
which is consistent with T-duality: the exponent is manifestly
invariant and the explicit $p^1$-dependent factor cancels against a
similar term in the measure factor, so that the resulting expression
depends only on the T-duality invariant quantities $q^2$ and $S$. This
confirms the importance of the measure factor $\sqrt{\Delta^-}$ in
(\ref{eq:laplace-osv}).

Because the real part of $S$ becomes large for large charges, we can
neglect the instanton contributions in $\Omega_k$ and use the
expression (\ref{eq:Omega-asymp}). This leads to, 
 \begin{eqnarray}
   \label{eq:perturbative-Omega}
   (S+\bar S)\partial_S \Omega_k &=& \ft12(S+\bar S)  -
   \frac{k+2}{4\,\pi}     \,,\nonumber \\
   (S+\bar S)^2 \partial_S \partial_{\bar S}\Omega_k &=&
   \frac{k+2}{4\,\pi}  \,,\nonumber \\
   (S+\bar S)^2 D_S\partial_S\Omega_k &=& (S+\bar S) -\frac{k+2}{4\,\pi}
   \,.
 \end{eqnarray}
Substituting these results into (\ref{eq:dets-electric}) one obtains, 
\begin{equation}
  \label{eq:Delta-electric-approx}
  \sqrt{\Delta^-(S, \bar S)} \propto (p^1)^{-2} (S+\bar S)^{(n+1)/2}
  \,\sqrt{\frac{k+2}{\pi}} \, \sqrt{S+\bar S - \frac{k+2}{2\pi}} \,,
\end{equation}
so that (\ref{eq:osv-electric}) acquires the form,
\begin{equation}
  \label{eq:osv-electric2}
  d(p^1,q_0,q_a) \propto  \int \frac{\mathrm{d}S\,\mathrm{d}\bar S}
  {(S+\bar S)^{k+4}}
  \, \sqrt{S+\bar S - \frac{k+2}{2\pi}} \;
  \exp\Big[ - \frac{\pi\,q^2}{S+\bar S} + 2\pi(S+\bar S)\Big] \,.
\end{equation}
Let us compare this result to the result obtained in
\cite{Dabholkar:2004yr,Dabholkar:2005x,Dabholkar:2005dt}, which is
also based on (\ref{eq:laplace-osv}) but without the integration
measure $\sqrt{\Delta^-}$. First of all, we note that
(\ref{eq:osv-electric2}) is manifestly invariant under T-duality, so
that no ad hoc normalization factor is needed.  Secondly, the above
result holds irrespective of the value of $n$, unlike in the
calculation without a measure, where one must choose the value
$n=2(k+2) -1$. Obviously the integral over the imaginary part of $S$
can be performed trivially and yields an overall constant. Upon
approximating the square root by $\sqrt{S+\bar S}$ the
resulting expression (\ref{eq:osv-electric2}) yields the following
semiclassical result for the entropy,
\begin{eqnarray}
  \label{eq:macroelec}
  S_{\mathrm{macro}} = \log d(q^2) = 4 \pi \,\sqrt{\vert q^2\vert/2} -
  \ft12 [(k+2)+ 1] \, \log |q^2| \;, 
\end{eqnarray}
which disagrees with the result (\ref{eq:microelec}) of microstate
counting. This is entirely due to the square root factor in the
integrand of (\ref{eq:osv-electric2}). As already noted in
\cite{Dabholkar:2005dt} there is a clear disagreement when the
instanton contributions are retained. We may also compare to
(\ref{eq:hetcalfelec}) that we shall derive later on the basis of the
mixed partition function, which is also in disagreement with the above
results. The situation is clearly unsatisfactory for small black
holes, in sharp contrast with the situation for large black holes
where there is a non-trivial agreement at the semiclassical level
between the various approaches.

In order to get a better handle on the subtleties in the electric
case, one may consider starting from (\ref{eq:laplace1}) and
integrating out the moduli fields $T^a$ in an exact manner, rather
than relying on (\ref{eq:laplace-osv}), which is based on a 
saddle-point approximation. Performing the
integral over the $T^a$ (which is Gaussian) and ignoring questions of
convergence, we obtain\footnote{There exists an analogous version of this
formula for the case of dyonic black holes.} 
\begin{eqnarray}
  \label{eq:the-integral}
  && 
 d(p^1,q_0,q_a) \propto  \int \,\frac{\mathrm{d}S\,\mathrm{d}\bar
 S}{(S+\bar S)^2}  \;  
 \frac{\mathrm{d}z\,\mathrm{d}\bar z\,\vert z\vert^{n+1}}{(1 -z -
 \bar z)^{(n-1)/2}} 
 \;   \mathrm{e}^{\pi\Sigma_{\mathrm{eff}}}   \nonumber \\[2mm]
&&\;\;   \times
 \Big\{ \frac{n(n-1)(z+\bar z)^2}{8\pi^2\vert
  z\vert^4(1-z-\bar z)^2} -   \frac{\vert (S+\bar S)^2\, D_S 
  \partial_S \Omega_k\vert^2}{\vert z\vert^4}  
  \nonumber \\[3mm]  
  && 
\qquad + \left[ \frac{(n-1)(z+\bar z)}{4\pi 
    \vert z\vert^2(1-z-\bar z)}    
  - \frac{(S+\bar S)\partial_S \Omega_k} {z^2} 
  - \frac{(S+\bar S)\partial_{\bar S}\Omega_k } {\bar z^2}     
  + \frac{(S+\bar S)^2 \partial_S {\bar
    \partial}_{\bar S} \Omega_k} {\vert z\vert^2}  \right]^2 
 \;  \Big\} \;, \nonumber\\
&&{~}  
\end{eqnarray}
where $\Sigma_{\rm eff}$ is given by 
\begin{eqnarray}
  \label{eq:Sigma-eff-elect}
\Sigma_{\mathrm{eff}} = - q^2 \, \frac{z+\bar z}{2(S + \bar S ) }
+ 4\Omega_k - \frac{2 (\bar z-1)}{z}
   (S+\bar S)  \partial_S\Omega_k
  - \frac{2(z-1)}{\bar z} (S+\bar S)
    \partial_{\bar S} \Omega_k \;,
\end{eqnarray}
and where $z$ is given by the S-duality invariant variable,
\begin{equation}
  \label{eq:def-z}
  z = \frac{Y^0\, (S+\bar S)} {\bar P(\bar S)} \;,
\end{equation}
where $P(S)$, defined in (\ref{eq:P-Q}), equals $p^1$ in the case at
hand. We observe that the integral (\ref{eq:the-integral}) is far more
complicated than the expression (\ref{eq:osv-electric}) resulting from
(\ref{eq:laplace-osv}).  In particular, we observe that the solution
for $Y^a$ induced by integrating out the $T^a$ reads
\begin{equation}
    \label{eq:Y-a}
    Y^a= - \,\frac{\eta^{ab} q_b}{2(S+\bar S)} \; z \,.
\end{equation}
Comparing with (\ref{eq:Ya-attractors}) shows that this only coincides
with the attractor value for $Y^a$ when $z=1$. Actually, the latter
equation is itself one of the attractor equations, as is clear from the first
equation in (\ref{eq:Ya-attractors}) . For this particular
value of $z$, $\pi\, \Sigma_{\rm eff}$ coincides with the exponent in
(\ref{eq:osv-electric}). Clearly, in order to better exhibit the 
difference between (\ref{eq:the-integral}) and
(\ref{eq:osv-electric}), it is crucial to perform the integral over
the variable $z$.  This, however, is a complicated integral.  On the
other hand, evaluating (\ref{eq:the-integral}) in saddle-point
approximation, neglecting as before the instanton contributions, gives
again the result (\ref{eq:macroelec}) while the saddle point
is still located at $z=1$. 


\section{More on heterotic black holes in $N=4$ compactifications}
\label{sec:heterotic-bh}
\setcounter{equation}{0}
In this section we will use the expressions for microscopic black hole
degeneracies to make contact with the macroscopic results described in
the previous sections.  Examples of these microscopic degeneracies are
provided by heterotic string theory compactified on a six-torus and by
the class of heterotic CHL models \cite{Chaudhuri:1995fk}. All these
models have $N=4$ supersymmetry.  A conjecture for the associated
microstate degeneracy has been put forward sometime ago in
\cite{Dijkgraaf:1996it} for the case of toroidally compactified
heterotic string theory, and more recently in \cite{Jatkar:2005bh} for
the case of CHL models.  In the toroidal case, the degeneracy is based
on the unique automorphic form $\Phi_{10}$ of weight $10$ under the
genus two modular group $\mathrm{Sp}(2,\mathbb{Z})$. This proposal has
recently received further support in \cite{Shih:2005uc}.  For the CHL
models, the degeneracy is based on the modular form $\Phi_k$ of weight
$k$ under a subgroup of the genus-two modular group
$\mathrm{Sp}(2,\mathbb{Z})$. It should be noted that the second
proposal is only applicable for states carrying electric charges
arising from the twisted sector \cite{Jatkar:2005bh}. 

It can be shown \cite{Cardoso:2004xf,Jatkar:2005bh} that, for large
charges, the asymptotic growth of the degeneracy of 1/4-BPS dyons in
the models discussed above precisely matches the macroscopic entropy
of dyonic black holes given in (\ref{eq:nonholoSigma}), with the
dilaton $S$ determined in terms of the charges through
(\ref{eq:attractor-S}).  This is reviewed in the next subsection.
\subsection{Asymptotic growth}
\label{sec:asymptotic-growth}
The degeneracy of $1/4$-BPS dyons in the class of models discussed
above, is captured by automorphic forms $\Phi_k (\rho,
\sigma,\upsilon)$ of weight $k$ under $\mathrm{Sp}(2,\mathbb{Z})$ or
an appropriate subgroup thereof \cite{Dijkgraaf:1996it,Jatkar:2005bh}.
The three modular parameters, $\rho,\sigma,\upsilon$, parametrize the
period matrix of an auxiliary genus-two Riemann surface which takes the
form of a complex, symmetric, two-by-two matrix. The case $k =10$
corresponds to $\Phi_{10}$, which is the relevant modular form for the
case of toroidal compactifications \cite{Dijkgraaf:1996it}.  The
microscopic degeneracy of $1/4$-BPS dyons in a given model takes the
form of an integral over an appropriate 3-cycle,
\begin{eqnarray}
   \label{eq:dphik}
 d_k(p,q) = \oint \de \rho\,\de\sigma\,\de\upsilon \; 
 \frac{{\rm e}^{\mathrm{i} \pi [
     \rho\, p^2 + \sigma\, q^2 + (2 \upsilon -1)\, p \cdot q]}}
 {\Phi_k(\rho,\sigma,\upsilon)} \;,
\end{eqnarray}
where we have included a shift of $\upsilon$, following
\cite{Shih:2005he}. It is important to note that the charges are in
general integer, with the exception of $q_1$ which equals a multiple
of $N$, and $p^1$ which is fractional and quantized in units of $1/N$.
Consequently $p^2/2$ and $ p \cdot q$ are quantized in integer units,
whereas $q^2/2$ is quantized in units of $1/N$. The inverse of the
modular form $\Phi_k$ takes the form of a Fourier sum with integer
powers of $\exp[2\pi\im \rho]$ and $\exp[2\pi\im \upsilon]$ and
fractional powers of $\exp[2\pi\im\sigma]$ which are multiples of
$1/N$. The 3-cycle is then defined by choosing integration contours
where the real parts of $\rho$ and
$\upsilon$ take values in the interval $(0,1)$ and the real part of
$\sigma$ takes values in the interval $(0,N)$. The precise definition
of $\Phi_k$ is subtle and we refer to \cite{Jatkar:2005bh} for further
details.

The formula (\ref{eq:dphik}) is invariant under both S-duality, which
is a subgroup of the full modular group, and T-duality.  Target-space
duality invariance is manifest, as the integrand only involves the
three T-duality invariant combination of the charges given in
(\ref{eq:chargeinvariants}).  To exhibit the invariance under
S-duality, one makes use of the transformation properties of the
charges as well as of the integration variables $\rho,
\sigma,\upsilon$. Since the result depends on the choice of an
integration contour, S-duality invariance is only formal at this
point.

The function $\Phi_k$ has zeros which induce corresponding poles in
the integrand whose residues will yield the microscopic degeneracy.
Since $\Phi_{k}$ has zeros in the interior of the Siegel half-space in
addition to the zeros at the cusps, the value of the integral
(\ref{eq:dphik}) depends on the choice of the integration 3-cycle. It
is possible to determine the poles of $\Phi_{k}^{-1}$ which are
responsible for the leading and subleading contributions to $d_k(p,q)$,
as was first shown in \cite{Cardoso:2004xf} for $k=10$, and
generalized recently to other values of $k$ in \cite{Jatkar:2005bh}.
Below we briefly summarize this result.

When performing an asymptotic evaluation of the integral
(\ref{eq:dphik}), one must specify which limit in the charges is
taken. `Large' black holes correspond to a limit where both electric
and magnetic charges are taken to be large.
More precisely, one takes $q^2 p^2 - (p \cdot q)^2 \gg 1$, and $q^2 +
p^2$ must be large and negative. This implies that the classical
entropy and area of the corresponding black holes are finite.  Under a
uniform scaling of the charges the dilaton will then remain finite; to
ensure that it is nevertheless large one must assume that $\vert
p^2\vert$ is sufficiently small as compared to $\sqrt{q^2p^2 -(p\cdot
  q)^2}$. In this way one can recover the nonperturbative string
corrections, as was stressed in \cite{Cardoso:2004xf}.

The leading behaviour of the dyonic degeneracy is associated with the
rational quadratic divisor ${\cal D} = \upsilon + \rho \sigma -
\upsilon^2=0$ of $\Phi_{k}$, near which $\Phi_k$ takes the form, 
\begin{eqnarray}
   \label{eq:Phik}
   \frac{1}{\Phi_k (\rho, \sigma,\upsilon)}  \approx
   \frac{1}{\mathcal{D}^2} \; \frac{1}
   {\sigma^{- (k +2)} \, f^{(k)} ( \gamma') \, f^{(k)} ( \sigma')} 
   +\mathcal{O}(\mathcal{D}^0) \;,
\end{eqnarray}
where
\begin{eqnarray}
   \label{eq:gsprime}
\gamma' = \frac{ \rho \sigma - \upsilon^2}{\sigma} \;\;\;,\;\;\;
 \sigma' = \frac{\rho \sigma - (\upsilon-1)^2}{\sigma} \;.
\end{eqnarray}
The cusp forms $f^{(k)}$ and their transformation rules have been
defined in (\ref{eq:Gamma-cusp}) and (\ref{eq:f-S-dual}),
respectively.  Here we note that \cite{Cardoso:2004xf} and
\cite{Jatkar:2005bh} differ from one another in the way the forms
$\Phi_k$ are expanded and in the expansion variables used.  This, for
instance, results in different definitions of $\sigma'$, which do,
however, agree on the divisor. Here we follow \cite{Jatkar:2005bh}.

Clearly, $\Phi_k$ has double zeros at $\upsilon_{\pm} = \ft12 \pm \ft12
\sqrt{1 +4 \rho \sigma}$ on the divisor.  The evaluation of
the integral (\ref{eq:dphik}) proceeds by first evaluating the contour
integral for $\upsilon$ around either one of the poles
$\upsilon_{\pm}$, and subsequently evaluating the two remaining
integrals over $\rho$ and $\sigma$ in saddle-point approximation.  The
saddle-point values of $\rho, \sigma$, and hence of $\upsilon_{\pm}$,
can be parametrized by
\begin{eqnarray}
\label{eq:rsvfixed}
\rho = \frac{\im |S|^2}{S + {\bar S}} \;, \qquad
 \sigma = \frac{\mathrm{i}}{S + \bar S}
\;,\qquad \upsilon_{\pm} = \frac{S}{S + \bar S} \;.
\end{eqnarray}
As argued in \cite{Cardoso:2004xf}, these values describe the unique
solution to the saddle-point equations for which the state degeneracy
$d_k(p,q)$ takes a real value. The resulting expression for $\log
d_k(p,q)$ precisely equals (\ref{eq:nonholoSigma}), with $S$ given by
the dilaton and expressed in terms of the charges through the
attractor equation (\ref{eq:attractor-S}). The result is valid up to a
constant and up to terms that are suppressed by inverse powers of the
charges.  Other divisors are expected to give rise to exponentially
suppressed corrections to the microscopic entropy $S_{\rm micro} =
\log d_k(p,q)$.

The asymptotic degeneracy can also be compared with the expression
(\ref{eq:complex-saddle}). However, we already argued that for large
black holes, the ratio of the two determinants equals one, up to
subleading terms that are inversely proportional to the charges.
Therefore, to this order of accuracy, the asymptotic degeneracy
computed from (\ref{eq:dphik}) is in precise agreement with
(\ref{eq:complex-saddle}), and hence correctly reproduced by the
proposal (\ref{eq:laplace1}).

\subsection{The mixed partition function}
\label{sec:next}
A more refined test of the proposal (\ref{eq:laplace1}) consists in
checking whether the mixed partition sum $Z_k(p,\phi)$ associated with
the microstate degeneracies $d_k(p,q)$ and defined by the first
equation of (\ref{eq:partition1-osv}), 
agrees with the
right-hand side of that same equation. The latter was derived from
(\ref{eq:laplace1}) by a saddle-point integration over the imaginary
part of the $Y^I$.  As discussed in
section~\ref{sec:partition-functions}, the right-hand side of
(\ref{eq:partition1-osv}) may require an explicit periodicity sum over
discrete imaginary shifts of the $\phi$.  Since the right-hand side of
(\ref{eq:partition1-osv}) results from a saddle-point evaluation, we
expect to find perturbative as well as non-perturbative corrections to
it. Both these features will show up in the examples discussed below.

In the following, we will first compute the mixed partition function
$Z_k(p,\phi)$ for $N=4$ dyons (corresponding to large black holes) in
the class of $N=4$ models discussed above.  For toroidally
compactified heterotic string theory, this mixed partition function
was recently computed in \cite{Shih:2005he} for the case when $p^0=0$.
A generalization to the case $p^0 \neq 0$ was reported in
\cite{Kappeli:2005nm}.  Here the same techniques are used to compute
the mixed partition function for CHL models.  Next, we compute the
(reduced) partition function for electrically charged 1/2-BPS states
(which correspond to small black holes) in the same class of models.

\subsubsection{Large (dyonic) black holes}
\label{sec:sec:next-dyonic} 
We start by noting that the microstate degeneracies must be consistent
with T-duality, so that the $d_k(p,q)$ can be expressed in terms of
the three invariants $Q\equiv q^2$, $P\equiv p^2$ and $R\equiv p\cdot
q$. Therefore we replace the sums over $q_0$ and $q_1$ in
(\ref{eq:partition1-osv}) by sums over the charges $Q$ and $P$,
related by the identities,\footnote{
  Observe that we will be assuming that both $p^0$ and $p^1$ are
  non-vanishing. When $p^0=0$, as was the case in \cite{Shih:2005he}, 
  the unrestricted sums can be replaced by sums over $Q$ and $R$.}
\begin{equation}
  \label{eq:Q-P-ids}
  q_0 = \frac1{2p^1} (Q+\ft12 q_a\eta^{ab}q_b) \,,\qquad    
  q_1 = - \frac1{2p^0} (P+2\, p^a\eta_{ab}p^b) \,.
\end{equation}
However, only those values of $Q$ and $P$ are admissible that lead to
integer-valued charges $q_0$ and charges $q_1$ that are a multiple
of $N$, also taking into account that $Q$ is quantized in units of
$2/N$ and that $P$ is even. These restrictions can be implemented by
inserting the series $L^{-1}\sum_{l=0}^{L-1} \;\exp[2\pi\im \,l\,
K/L]$, where $K$ and $L$ are integers (with $L$ positive), which
projects onto all integer values for $K/L$. The use of this formula
leads to the following expression, 
\begin{eqnarray}
\label{zQP}
  Z_k(p,\phi) &=& \frac{1}{N^2 p^0 p^1 }\sum_{
    \begin{array}{l} \scriptstyle
      \phi^0  \rightarrow\phi^0+ 2\im\, l^0\\[-.2mm]
      \scriptstyle 
      \phi^1 \rightarrow\phi^1+ 2\im \,l^1/N \end{array} }
  \sum_{q_a, Q,P}  \, d_k(Q,P,R)  \nonumber\\
  &&\times \exp \left[\frac{\pi \phi^0}{2p^1}(Q + \ft12 q_a \eta^{ab}
  q_b)  - \frac{ \pi \phi^1}{2p^0}
  (P + 2 p^a \eta_{ab} p^b) + \pi q_a \phi^a
  \right] \;,
\end{eqnarray}
with $R$ given by
\begin{equation}
  R =  \frac{p^0}{2p^1}(Q + \ft12 q_a \eta^{ab} q_b)+
  \frac{p^1}{2p^0}(P + 2 p^a \eta_{ab} p^b) + q_a p^a \,.
\end{equation}
In (\ref{zQP}) the summation over imaginary shifts of $\phi^0$ and
$\phi^1$ is implemented by first replacing $\phi^{0}
\rightarrow\phi^{0}+ 2 \im l^{0}$ and $\phi^{1} \rightarrow\phi^{1}+ 2
\im l^{1}/N$ in each summand, and subsequently summing over the
integers $l^0 = 0,\ldots, N p^1-1$ and $l^1 = 0,\ldots,Np^0-1$. The
sums over $l^{0,1}$ enforce that only those summands, for which $(Q +
\ft12 q_a \eta^{ab} q_b )/2p^1$ is an integer and $(P + 2 p^a
\eta_{ab} p^b)/2p^0$ is a multiple of $N$, give a non-vanishing 
contribution to $Z_k(p,\phi)$. 

Next, we perform the sums over $Q$ and $P$ without any restriction,
using (\ref{eq:dphik}) and taking into account that $N Q/2$ and $P/2$
are integer valued. Provided we make a suitable choice for the
integration contours for $\sigma$ and $\rho$, both sums can be
rewritten as sums of delta-functions, which imply that $\sigma$ and $\rho$ 
are equal to $\sigma(\upsilon)$ and 
$\rho(\upsilon)$, up to
certain integers,  where 
\begin{equation}
\label{eq:sigmarho}
  \begin{split}
\sigma(\upsilon)&= -\frac{\phi^0}{2\im p^1} - (2\upsilon-1)
   \frac{p^0}{2p^1}\,,\\
   \rho(\upsilon)&= \frac{\phi^1}{2\im p^0} - (2\upsilon-1)
   \frac{p^1}{2p^0}\,. \\
     \end{split}
\end{equation}
The required choice for the integration contours implies
$\mbox{Im}\,\sigma = \mbox{Im} \, \sigma(\upsilon)$ and
$\mbox{Im} \, \rho = \mbox{Im} \, \rho(\upsilon)$,
for given
$\upsilon$. The sums of delta-functions then take the form
$\sum_{n\in\mathbb{Z}} N \,
\delta(\mbox{Re} \, \sigma - \mbox{Re} \, \sigma(\upsilon)- n N)$ and
$\sum_{m\in\mathbb{Z}} \delta(\mbox{Re} \, \rho - \mbox{Re} \, 
\rho (\upsilon) -
m)$, respectively. Note that the shifts in the arguments of the
delta-functions are precisely generated by additional shifts of
$\phi^0$ and $\phi^1$,
\begin{equation}
  \label{eq:big-shifts}
  \phi^0\to \phi^0 + 2\im p^1 N\,n\,,\qquad 
  \phi^1\to \phi^1 + 2\im p^0 \,m\,.
\end{equation}
However, the resulting integrals do not depend on these shifts as they
turn out to be periodic under (\ref{eq:big-shifts}). Therefore only one
of the delta-function contributes, so that the integrations over
$\sigma$ and $\rho$ result in 
\begin{equation}
  \label{mpfpk}
\begin{split}
    Z_k(p,\phi) &= \frac{1}{N\, p^0  p^1}
    \sum_{ \begin{array}{l} \scriptstyle
      \phi^0  \rightarrow\phi^0+ 2\im\, l^0\\[-.2mm] 
      \scriptstyle 
      \phi^1 \rightarrow\phi^1+ 2\im \,l^1/N \end{array} }
  \sum_{q_a}\; 
  \int\,\de\upsilon\;\frac{1}
  {\Phi_{k}(\rho(\upsilon),\sigma(\upsilon),\upsilon)}\\
  &
  \times \exp\left(- \im \pi \left[ \ft12 \sigma(\upsilon) \,
      q_a \eta^{ab} q_b
      +2\rho(\upsilon)\, p^a \eta_{ab} p^b + \im q_a \left( \phi^a +
      \im (2\upsilon-1) p^a\right) 
    \right] \right) \,.
\end{split}
\end{equation}
Since the integrand is invariant under the shifts
(\ref{eq:big-shifts}),  the explicit sum over shifts 
with $l^0 = 0, \ldots, Np^1 -1$ and $l^1 = 
0, \ldots, Np^0-1$ ensures that the  partition function 
(\ref{mpfpk}) is invariant under shifts
 $\phi^a
\rightarrow \phi^a + 2 \mathrm{i} $, as well as under shifts $\phi^{0}
\rightarrow\phi^{0}+ 2 \im$ and $\phi^{1} \rightarrow\phi^{1}+ 2
\im/N$.

Subsequently we perform a formal Poisson resummation of the charges
$q_a$, i.e., we ignore the fact that $\eta_{ab}$ is not positive
definite. We obtain (up to an overall numerical constant), 
\begin{eqnarray}
\label{eq:partpoisson}
   Z_k(p,\phi) &=& \frac{\sqrt{\det[-\eta_{ab}]}}
   {N\, p^0 p^1} \,  \,
     \sum_{\mathrm{shifts}}  
     \int \,\de\upsilon \; \frac{1}
       {\sigma (\upsilon)^{(n-1)/2}\; \Phi_k (\rho(\upsilon), \sigma
       (\upsilon), \upsilon)} \nonumber\\
     && \times \exp \left( - \im \pi \left[  2 p^a
         \eta_{ab} p^b \, \rho(\upsilon) 
         + \frac{ ( \phi^a +\im (2\upsilon-1)p^a) \, \eta_{ab} \,
           ( \phi^b +\im(2\upsilon-1)p^b)}
         {2 \sigma (\upsilon)} \right]\right) \;, \nonumber\\
     &&{~} 
 \end{eqnarray}
where here and henceforth the sum over shifts denotes the 
infinite sum over shifts $\phi^a \rightarrow \phi^a + 2\im$, 
together with the finite sums over shifts in $\phi^0$ and 
$\phi^1$. This result is completely in line with what was discussed
below (\ref{eq:partition-hesse}). 

Now we perform the contour integral over $\upsilon$. This integration
picks up the contributions from the residues at the various poles of
the integrand. The leading contribution to this sum of residues stems
from the zeros of $\Phi_k$.  For large magnetic charges $p$ and large
scalars $\phi$, the leading contribution to the mixed partition
function $Z_k(p, \phi)$ is expected to be associated with the rational
quadratic divisor ${\cal D} = \upsilon + \rho \sigma - \upsilon^2=0$
of $\Phi_{k}$, near which $\Phi_k$ takes the form (\ref{eq:Phik})
\cite{Cardoso:2004xf,Jatkar:2005bh}. This is the divisor
responsible for the leading contribution to the entropy
\cite{Dijkgraaf:1996it,Cardoso:2004xf,Jatkar:2005bh}, and hence it is
natural to expect that this divisor also gives rise to the leading
contribution to the free energy, and therefore to the mixed partition
function.  Then, other poles of the integrand in
(\ref{eq:partpoisson}) give rise to exponentially suppressed
contributions.

Inserting $\rho(\upsilon)$ and $\sigma(\upsilon)$ into ${\cal D}$
yields
\begin{equation}
\label{eq:D-dyonic}
  {\cal D}= 2(\upsilon-\upsilon_*) \,\frac{\phi^0 p^1 - \phi^1 p^0}
  {4\im p^0p^1}\,,
\end{equation}
with $\upsilon_*$ given by 
\begin{equation}
\label{eq:v-star}
2\upsilon_* = 1-\im\, 
\frac{\phi^0\phi^1+p^1p^0}{\phi^0p^1-\phi^1p^0}\,.
\end{equation}
We observe that the quadratic piece in $\upsilon$ has canceled, and
that ${\cal D}$ has therefore a simple zero. Performing the contour
integral over $\upsilon$ then yields (again, up to an overall
numerical constant),
\begin{equation}
  \label{eq:zres}
  \begin{split}
    Z_k(p,\phi) &= \frac{p^0 p^1\,\sqrt{\det[- \eta_{ab}]}}{N}  
      \sum_{\mathrm{shifts}}\; \frac{1}{(\phi^0p^1-\phi^1p^0)^2}
      \\[2mm] 
  &\times \frac{\de}{\de\upsilon}\left[
  \frac{
    \exp \left( - \im \pi \left[  2 p^a
        \eta_{ab} p^b \, \rho(\upsilon) + \frac{ ( \phi^a +\im
          (2\upsilon-1)p^a) \, \eta_{ab} \,( \phi^b +\im
          (2\upsilon-1)p^b)}{2 \sigma (\upsilon)}  \right] \right)}
          {\sigma(\upsilon)^{(n-2k-5)/2}\;f^{(k)}(\gamma'(\upsilon))\; 
          f^{(k)}(\sigma'(\upsilon))}\right]_{\upsilon = \upsilon_*},
 \end{split}
\end{equation}
where we made use of (\ref{eq:Phik}) and we discarded the
exponentially suppressed contributions originating from other possible
poles of the integrand.

Using (\ref{eq:sigmarho}) we can determine the following expressions
for the derivatives with respect to $\upsilon$ on the divisor,
\begin{equation}
  \label{eq:deriv}
  \begin{array}{rl}
  \displaystyle 
  \frac{\de\rho (\upsilon)}{\de \upsilon}\Big|_{\upsilon=\upsilon_*} 
  &   \displaystyle
  =   - \frac{p^1}{p^0} \;, \\[3mm]
  \displaystyle
  \frac{\de\sigma (\upsilon)}{\de\upsilon}\Big|_{\upsilon=\upsilon_*} 
  &  \displaystyle
  =  - \frac{p^0}{p^1} \;,  
  \end{array}
  \qquad
  \begin{array}{rl}
  \displaystyle
  \frac{\de \gamma'(\upsilon)}{\de\upsilon}\Big|_{\upsilon=\upsilon_*} 
  &  \displaystyle
  = - \frac{[p^1\sigma +p^0 v]^2}{p^0 p^1\,\sigma^2} \;,  \\[3mm]
  \displaystyle
  \frac{\de\sigma'(\upsilon)}{\de\upsilon}\Big|_{\upsilon=\upsilon_*}
  &  \displaystyle
  =  - \frac{[p^1\sigma+p^0(v-1)]^2}{p^0 p^1\,\sigma^2}\;, 
  \end{array}
\end{equation}
where $\upsilon, \rho$ and $\sigma$ on the right-hand side refer to the
values of these variables on the divisor, i.e., $\upsilon_*$,
$\rho_*=\rho(\upsilon_*)$ and $\sigma_*=\sigma(\upsilon_*)$.
Inserting the above expressions into (\ref{eq:zres}), we obtain,
\begin{equation}
  \label{eq:zres2}
  \begin{split}
    Z_k(p,\phi) &= \frac{\sqrt{\det[- \eta_{ab}]}}{N}
    \sum_{\mathrm{shifts}}\; \frac{\sigma^{-(n+3)/2}}
    {(\phi^0p^1-\phi^1p^0)^2}
    \\[2mm]
    &\times \exp\Big(-\frac{\im\pi}{2\sigma} \left[\phi^a\eta_{ab}
        \phi^b - p^a\eta_{ab}p^b + 2\im\,p^a\eta_{ab}\phi^b
        (2\upsilon-1) \right] \\ 
    &\hspace{14mm}
      - \ln f^{(k)}(-\upsilon/\sigma) - \ln
        f^{(k)}((\upsilon-1)/\sigma)  + (k+2)\ln \sigma \Big) \\[2mm]
    &\times \Big[- \ft12\im\pi (p^a\phi^0 - p^0 \phi^a) \eta_{ab}
    (p^b\phi^0 - p^0 \phi^b) + \ft12 (n-2k-5) (p^0)^2 \sigma \\
    &\hspace{5mm}
    + [p^1\sigma +p^0 v]^2\,
    \big[\ln f^{(k)}(-\upsilon/\sigma)\big]^\prime  +
    [p^1\sigma+p^0(v-1)]^2\, 
    \big[\ln f^{(k)}((\upsilon-1)/\sigma)\big]^\prime\Big] \;. 
 \end{split}
\end{equation}

To make contact with the macroscopic expressions, we define $S$ and
$\bar S$ in terms of $\phi^0$ and $\phi^1$, according to
(\ref{eq:dilaton}). Substituting the corresponding expressions into
(\ref{eq:sigmarho}) and (\ref{eq:v-star}), we recover precisely the
expressions for the divisor values of $\rho,\sigma,\upsilon$ in terms
of $S$ and $\bar S$ that were shown in (\ref{eq:rsvfixed}). Likewise
we use $T^a = ( - \im \phi^a + p^a)/(\phi^0 + \im p^0)$ and ${\bar
  T}^a = (\im\phi^a + p^a)/(\phi^0 - \im p^0)$. Note that under the
periodicity shifts, $S, \bar S, T^a,\bar T^a$ should be treated as
functions of $\phi^I$. After being subjected to such a shift, $S$ and
$\bar S$, and $T^a$ and $\bar T^a$, respectively, are no longer
related by complex conjugation.

We now note that (\ref{eq:zres2}) takes the same form as
(\ref{eq:partition1-osv}). The exponential factor in (\ref{eq:zres2})
coincides precisely with $\exp[\pi{\mathcal F}_{\mathrm{E}}(p,\phi)]$
after substituting (\ref{eq:rsvfixed}), so that the prefactor
$\sqrt{\Delta^-}$ should be identified with the remaining terms. Hence
we obtain (up to an overall numerical constant), 
\begin{equation}
  \label{eq:delta-compare}
  \sqrt{\Delta^-} = \frac{(S + \bar S)^{(n-3)/2}\sqrt{\det[-
  \eta_{ab}]}}{2\,|Y^0|^2} \; 
  \left( {\hat K} + 4 (S + {\bar S})^2 \partial_S \partial_{\bar S} \Omega
  + \frac{(n-1)}{4 \pi} \frac{(Y^0 - {\bar Y}^0)^2}{|Y^0|^2}\right),
\end{equation} 
where we also used that $Y^0= (\phi^0+\im p^0)/2$. The above results
for the mixed partition function of $N=4$ dyons (with generic charges)
in CHL models is exact, up to exponentially suppressed corrections.
When setting $p^0=0$ the resulting expression for the toroidal case
($k=10$) agrees with the one found in \cite{Shih:2005he},
up to a subtlety involving the periodicity sums.  The
expression (\ref{eq:delta-compare}) is consistent with our previous
result (\ref{eq:dets-general}) in the limit of large charges. In that
limit the term proportional to $\hat K$ dominates, as we explained in
section~\ref{sec:partition-functions}. Recall, however, that
$\sqrt{\Delta^-(p,\phi)}$ enters into (\ref{eq:partition1-osv}) in the
context of a saddle-point approximation, which is expected to be
subject to further perturbative and non-perturbative corrections.

There is, however, an issue with regard to the number of moduli, which
depends on the integer $n$. In the context of the above calculation,
$n+1$ equals the rank of the gauge group of the corresponding CHL
model. On the other hand, in the context of $N=2$ supersymmetry $n$
defines the number of matter vector supermultiplets. In this case the
rank of the gauge group is still also equal to $n+1$. However, the
value taken for $n$ in the case of $N=4$ supersymmetry differs from
the $N=2$ value by four. The difference is related to the six
graviphotons of pure $N=4$ supergravity, whose $N=2$ decomposition is
as follows.  One graviphoton belongs to the $N=2$ graviton multiplet,
another graviphoton belongs to an $N=2$ vector multiplet, whereas the
four remaining graviphotons belong to two $N=2$ gravitino supermultiplets. 
When the description of the $N=4$ supersymmetric black holes is based
on $N=2$ supergravity, the above results seem to indicate that the
charges (and the corresponding electrostatic potentials $\phi$)
associated with the extra gravitini should be taken into
account. This question is particularly pressing for the small
(electric) $N=4$ black holes, where saddle-point approximations are
more cumbersome. For that reason we will briefly reconsider the mixed
partition function for the case of small black holes in the next
subsection.

\subsubsection{Small (electric) black holes}
\label{sec:sec:next-electric}
Here, we compute the partition function
for electrically charged 1/2-BPS states in CHL models.
These are states with vanishing charges $q_1, p^0$ and $p^a$.  We therefore
consider the reduced partition sum, 
\begin{eqnarray}
   \label{eq:mixedrpf}
   Z_{\rm R}(p,\phi) = \sum_{q_0,q_a} d(q^2) \; {\rm e}^{\pi [
     q_0\phi^0 + q_a \phi^a]} \;.
\end{eqnarray}
As in the previous subsection, we replace the sum over the charges
$q_0$ by a sum over the charges $Q \equiv q^2$, where we recall that
$Q$ is quantized in units of $2/N$ \cite{Jatkar:2005bh}. Following the
same step as in the derivation of (\ref{zQP}), this results in
\begin{eqnarray}
  \label{zQ}
  Z_{\rm R}(p,\phi) = \frac{1}{N\, p^1 } 
  \sum_{\phi^{0}\rightarrow\phi^{0}+ 2\im l^{0}} \sum_{q_a,Q} 
  \, d(Q)\; \exp \left[
  \frac{\pi \phi^0}{2p^1}(Q + \ft12 q_a \eta^{ab} q_b)+ \pi q_a \phi^a
  \right]\,.
\end{eqnarray}
Here, the integers $l^0$ run over $l^0 = 0,\ldots, N p^1-1$.  

Next, we perform the sum over $Q$ by using the integral expression
for the electric degeneracies \cite{Jatkar:2005bh},
\begin{equation}
   \label{eq:electdeg}
   d(Q) = \oint \de \sigma \;\sigma^{k+2}\, \frac{{\rm e}^{\mathrm{i}
   \pi \sigma Q}}{f^{(k)}(-1/\sigma)}  \;,
\end{equation}
where $\sigma$ runs in the strip $\sigma \sim \sigma +N$.  Observe
that (\ref{eq:electdeg}) has the asymptotic expansion, 
\begin{eqnarray}
  \label{eq:microelec}
  \log d (Q) = 4 \pi \,\sqrt{ \ft12
    |Q| } - \ft12 [(k+2)+ \ft32 ] \, \log |Q| \;.
\end{eqnarray}
By making a suitable choice for the integration contour of $\sigma$,
the sum over $Q$ can be rewritten as a sum over
delta-functions, $\sum_{n\in\mathbb{Z}} N \, \delta({\rm Re} \, \sigma
- {\rm Re} \, \sigma_*-
n N)$, where $\sigma_* = - \phi^0/(2 \im p^1)$. The sum over $n$ is
generated by imaginary shifts of $\phi^0$ according to 
\begin{equation}
  \label{eq:elec-shift}
  \phi^0 \to \phi^0 + 2\im p^1 N\,n\,,
\end{equation}
just as before. 
However, the resulting integral does not depend on these shifts as 
it is periodic under (\ref{eq:elec-shift}). Therefore only one
of the delta-function contributes, so that the integration over
$\sigma$ results in 
\begin{eqnarray}
  \label{eq:Z-next} 
    Z_{\rm R}(p,\phi) = \frac{1}{p^1}\sum_{\phi^{0}\rightarrow\phi^{0}+
    2 \im l^{0}} \, \frac{\sigma_*^{k+2}} { f^{(k)} (-1/\sigma_*)} 
  \,\sum_{q_a} \, \exp \left( \im \pi \left[ -\ft12 \sigma_* \, q_a
      \eta^{ab} q_b - \im q_a \,\phi^a \right] \right) .
\end{eqnarray}
Since (\ref{eq:Z-next}) is invariant under the shifts
(\ref{eq:elec-shift}),  the explicit sum over shifts 
with $l^0 = 0, \ldots, Np^1 -1$ 
ensures that the reduced partition function 
(\ref{eq:Z-next}) is invariant under shifts
 $\phi^a
\rightarrow \phi^a + 2 \mathrm{i} $ as well as under shifts $\phi^{0}
\rightarrow\phi^{0}+ 2 \im$.

The next step is to perform a Poisson resummation of the charges
$q_a$, ignoring, as before, that $\eta_{ab}$ is not positive
definite. This yields (up to an overall numerical constant), 
\begin{eqnarray}
  \label{eq:electricpar}
   Z_{\rm R} (p,\phi) = \frac{\sqrt{\det[-\eta_{ab}]}}{p^1}\,     
     \sum_{\phi^{0,a}\rightarrow\phi^{0,a}+
    2 \im l^{0,a}} \frac{\sigma_*^{- (n-1)/2 + k + 2 }} 
   {f^{(k)} (-1/\sigma_*)}\, \exp \left( - \im \pi
    \frac{ \phi^a \, \eta_{ab} \, \phi^b }{2 \sigma_*}\right)\;,
\end{eqnarray}
where $l^a \in \mathbb{Z}$.

Now we recast (\ref{eq:electricpar}) in terms of the scalar field $S$
given in (\ref{eq:dilaton}).  Because $p^0 =0$ in the electric case,
$S= (p^1 -\im \phi^1)/\phi^0$ so that $\sigma_* = \im/ (S + {\bar
  S})$, precisely as in (\ref{eq:rsvfixed}). Clearly, the result
(\ref{eq:electricpar}) can now be factorized as follows (again up to
an overall numerical constant),
\begin{equation}
  \label{eq:Delta-F}
  Z_{\rm R}(p,\phi) = \sum_{\mathrm{shifts}}\,
  \sqrt{\tilde\Delta^-(p,\phi)} \; \ee^{\pi 
  \,\tilde{\mathcal{F}}_{\mathrm{E}}(p,\phi)}\;, 
\end{equation}
where 
\begin{eqnarray}
  \label{eq:hetcalfelec}
  {\tilde {\cal F}}_{\rm E}(p,\phi) &=& - \ft12(S + {\bar S})
  \phi^a\eta_{ab} \phi^b - \frac{(k+2)}{\pi} \log (S + {\bar S})
  - \frac{1}{\pi} \log f^{(k)} ( S + {\bar S}) \;, \nonumber\\
  \tilde\Delta^-(p,\phi)&=&
  \frac{\det[-\eta_{ab}]}{(p^1)^2}\, (S + \bar S)^{n-1} \;. 
\end{eqnarray}
Although (\ref{eq:Delta-F}) is of the same form as the right-hand side
of (\ref{eq:partition1-osv}), the quantities ${\tilde {\cal F}}_{\rm
  E}(p,\phi)$ and $\tilde\Delta^-(p,\phi)$ do not at all agree with
(\ref{eq:hetcalf}) and (\ref{eq:dets-electric}). One of the most
conspicuous features is the fact that the partition function does not
depend on $\phi^1$, which is proportional to the imaginary part of
$S$. This is the result of the fact that for the electric black hole
we took $q_1=0$. This is undoubtedly related to the electric/magnetic
duality basis that has to be used here \cite{Ceresole,DKLL}. When
suppressing the instanton corrections, both (\ref{eq:hetcalf}) and
(\ref{eq:dets-electric}) also become functions of the real part of
$S$. In that case, ${\tilde {\cal F}}_{\rm E}(p,\phi)$ does coincide
with (\ref{eq:hetcalf}), but there is no way to reconcile
$\tilde\Delta^-(p,\phi)$ with (\ref{eq:Delta-electric-approx}).

\subsection*{Acknowledgements}
We wish to thank V. Cort\'es, A. Dabholkar, J. de Boer, F. Denef, G.
Gibbons, R. Kallosh, B. Pioline, A. Sen, D. Shih and E. Verlinde for
valuable discussions.  This work is partly supported by EU contracts
MRTN-CT-2004-005104 and MRTN-CT-2004-512194, and partly supported by
INTAS contract 03-51-6346. The work of T.M. was supported by the DFG
``Schwerpunktprogramm String\-theorie''.

\begin{thebibliography}{10}
\bibitem{Strominger:1996sh}
A.~Strominger and C.~Vafa, ``Microscopic origin of the {B}ekenstein-{H}awking
  entropy'', {\em Phys. Lett.} {\bf B379} (1996) 99--104,
\href{http://www.arXiv.org/abs/hep-th/9601029}{{\tt hep-th/9601029}}.
%
\bibitem{Maldacena:1997de}
J.M. Maldacena, A.~Strominger and E.~Witten, ``Black hole entropy in
  {M}-theory'', {\em JHEP} {\bf 12} (1997) 002,
\href{http://www.arXiv.org/abs/hep-th/9711053}{{\tt hep-th/9711053}}.
%
\bibitem{Vafa:1997gr}
C.~Vafa, ``Black holes and {C}alabi-{Y}au threefolds'', {\em Adv. Theor. Math.
  Phys.} {\bf 2} (1998) 207--218,
\href{http://www.arXiv.org/abs/hep-th/9711067}{{\tt hep-th/9711067}}.
%
\bibitem{LopesCardoso:1998wt}
G.L. Cardoso, B.~de~Wit and T.~Mohaupt, ``Corrections to macroscopic
  supersymmetric black-hole entropy'', {\em Phys. Lett.} {\bf B451} (1999)
  309--316,
\href{http://www.arXiv.org/abs/hep-th/9812082}{{\tt hep-th/9812082}}.
%
\bibitem{Wald:1993nt} R.M. Wald, ``Black hole entropy entropy is
  Noether charge'', {\it Phys. Rev.} {\bf D48} (1993) 3427, {\tt
    gr-qc/9307038}; V. Iyer and R.M. Wald, ``Some properties of
  Noether charge and a proposal for dynamical black hole entropy'',
  {\it Phys. Rev.}  {\bf D50} (1994) 846, {\tt gr-qc/9403028};
  T. Jacobson, G. Kang and R.C. Myers, ``Black hole entropy in higher
  curvature gravity'', {\it Phys. Rev.} {\bf D49} (1994) 6587, {\tt
    gr-qc/9312023}.
%
\bibitem{LopesCardoso:2000qm}
G.L. Cardoso, B.~de~Wit, J.~{K\"appeli} and T.~Mohaupt, ``Stationary {BPS}
  solutions in $N = 2$ supergravity with {$R^2$}-interactions'', {\em JHEP}
  {\bf 12} (2000) 019,
\href{http://www.arXiv.org/abs/hep-th/0009234}{{\tt hep-th/0009234}}.
%
\bibitem{Ooguri:2004zv}
H.~Ooguri, A.~Strominger and C.~Vafa, ``Black hole attractors and the
  topological string'', {\em Phys. Rev.} {\bf D70} (2004) 106007,
\href{http://www.arXiv.org/abs/hep-th/0405146}{{\tt hep-th/0405146}}.
%
\bibitem{Bershadsky:1993ta} M. Bershadsky, S. Cecotti, H. Ooguri and C. Vafa,
``Holomorphic anomalies in topological field theories'',
{\em Nucl. Phys.} {\bf B405} (1993) 279-304, {\tt hep-th/9302103};
``Kodaira-Spencer theory of gravity and exact
results for quantum string amplitudes'', {\em Commun. Math. Phys.}
{\bf 165} (1994) 311-428,
{\tt hep-th/9309140}.
%
\bibitem{Antoniadis:1994}
  I.~Antoniadis, E.~Gava, K.S.~Narain and T.R.~Taylor,
  ``Topological amplitudes in string theory'', 
  {\em Nucl. Phys.} {\bf B413} (1994) 162, 
  {\tt hep-th/9307158}.
%
\bibitem{Vafa:2004qa}
  C.~Vafa,
  ``Two dimensional Yang-Mills, black holes and topological strings,''
  {\tt hep-th/0406058}.
%
\bibitem{Aganagic:2004js}
  M.~Aganagic, H.~Ooguri, N.~Saulina and C.~Vafa,
  ``Black holes, q-deformed 2d {Yang-Mills}, and non-perturbative topological
  strings,''
  {\em Nucl. Phys.} {\bf B715} (2005) 304,
  {\tt hep-th/0411280}.
%
\bibitem{Caporaso:2005np}
N. Caporaso, M. Cirafici, L. Griguolo, S. Pasquetti, D. Seminara 
and R.J. Szabo,
``Black-holes, topological strings and large N phase transitions'',
{\tt hep-th/0512213}.
%
\bibitem{Aganagic:2005wn}
M. Aganagic, D. Jafferis and N. Saulina,
``Branes, Black Holes and Topological Strings on Toric Calabi-Yau Manifolds'',
{\tt hep-th/0512245}.
%
\bibitem{Gerasimov:2004yx}
  A.A.~Gerasimov and S.L.~Shatashvili,
  ``Towards integrability of topological strings. I: Three-forms on Calabi-Yau
  manifolds''.   {\em JHEP} {\bf 0411} (2004) 074,
  {\tt hep-th/0409238}.
%
\bibitem{Verlinde:2004ck}
E.~Verlinde, ``Attractors and the holomorphic anomaly'',
{\tt hep-th/0412139}.
%
\bibitem{Ooguri:2005vr}
  H.~Ooguri, C.~Vafa and E.~P.~Verlinde,
  ``Hartle-Hawking wave-function for flux compactifications'',
  {\tt hep-th/0502211}.
%
\bibitem{Dabholkar:2004yr}
A.~Dabholkar, ``Exact counting of black hole microstates'',
{\em Phys. Rev. Lett.} {\bf 94} (2005) 241301,
\href{http://www.arXiv.org/abs/hep-th/0409148}{{\tt hep-th/0409148}}.
%
\bibitem{Dabholkar:2005x} A.~Dabholkar, F. Denef, G.W. Moore and B. Pioline,
  ``Exact and asymptotic degeneracies of small black holes'',
{\em JHEP} {\bf 0508} (2005) 021,
  \href{http://www.arXiv.org/abs/hep-th/0502157}{{\tt
  hep-th/0502157}}.
%
\bibitem{Sen:2005ch} A. Sen, ``Black holes and the spectrum of half-BPS
States in {$N=4$} supersymmetric string theory'',
{\tt hep-th/0504005}.
%
\bibitem{Dabholkar:2005dt} A.~Dabholkar, F. Denef, G.W. Moore and
B. Pioline, 
``Precision counting of small black holes'',
{\em JHEP} {\bf 0510} (2005) 096, {\tt hep-th/0507014}.
%
\bibitem{Kappeli:2005nm}
  J.~K\"appeli,
  ``Partition function of dyonic black holes in {$N = 4$} string
  theory'', 
  {\tt hep-th/0511221}.

\bibitem{deWit:2005ya}
  B.~de Wit, 
  ``Supersymmetric black holes'', {\it Fortschr. Phys.} {\bf 54}
  (2006) 183, {\tt hep-th/0511261}.
%
\bibitem{Mohaupt:2005jd}
T. Mohaupt, ``Strings, higher curvature corrections and black holes",
\href{http://www.arXiv.org/abs/hep-th/0512048}{{\tt hep-th/0512048}}.
%



\bibitem{Ferrara:1995ih}
S.~Ferrara, R.~Kallosh and A.~Strominger, ``$N = 2$ extremal black holes'',
  {\em Phys. Rev.} {\bf D52} (1995) 5412--5416,
\href{http://www.arXiv.org/abs/hep-th/9508072}{{\tt hep-th/9508072}}.
%
\bibitem{Strominger:1996kf}
A.~Strominger, ``Macroscopic {entropy} of $N = 2$ {extremal} {black} {h}oles'',
  {\em Phys. Lett.} {\bf B383} (1996) 39--43,
\href{http://www.arXiv.org/abs/hep-th/9602111}{{\tt hep-th/9602111}}.
%
\bibitem{Ferrara:1996dd}
S.~Ferrara and R.~Kallosh, ``Supersymmetry and {a}ttractors'', {\em Phys. Rev.}
  {\bf D54} (1996) 1514--1524,
\href{http://www.arXiv.org/abs/hep-th/9602136}{{\tt hep-th/9602136}}.
%
%
\bibitem{Gibbons}
G.W. Gibbons, unpublished notes.
%
\bibitem{Ferrara:1997tw}
S. Ferrara, G.W. Gibbons and R. Kallosh,
``Black holes and critical points in moduli space'',
{\em Nucl. Phys.} {\bf B500} (1997) 75-93,
\href{http://www.arXiv.org/abs/hep-th/9702103}{{\tt hep-th/9702103}}.
%
\bibitem{Sen:2005wa}
A. Sen, ``Black hole entropy function and the attractor mechanism in
higher derivative gravity'', {\em JHEP} {\bf 0509} (2005) 038,
\href{http://www.arXiv.org/abs/hep-th/0506177}{{\tt hep-th/0506177}}.
%
\bibitem{Goldstein:2005hq}
K. Goldstein, N. Iizuka, R.P. Jena and S.P. Trivedi,
``Non-supersymmetric attractors'',
\href{http://www.arXiv.org/abs/hep-th/0507096}{{\tt hep-th/0507096}}.
%
\bibitem{Sen:2005iz}
A. Sen, ``Entropy function for heterotic black holes'',
\href{http://www.arXiv.org/abs/hep-th/0508042}{{\tt hep-th/0508042}}.
%
\bibitem{Kallosh:2005ax}
R. Kallosh, 
``New attractors'',
\href{http://www.arXiv.org/abs/hep-th/0510024}{{\tt hep-th/0510024}}.
%
\bibitem{Tripathy:2005qp}
 P.K. Tripathy and S.P. Trivedi, ``Non-supersymmetric attractors in
string theory'',
\href{http://www.arXiv.org/abs/hep-th/0511117}{{\tt hep-th/0511117}}.
%
\bibitem{Giryavets:2005nf}
A. Giryavets, ``New attractors and area codes'',
\href{http://www.arXiv.org/abs/hep-th/0511215}{{\tt hep-th/0511215}}.
%
\bibitem{Prester:2005qs}
P. Prester,
``Lovelock type gravity and small black holes in heterotic string theory'',
{\tt hep-th/0511306}.
%
\bibitem{Goldstein:2005rr}
K. Goldstein, R.P. Jena, G. Mandal and S.P. Trivedi,
``A C-function for non-supersymmetric attractors'', 
\href{http://www.arXiv.org/abs/hep-th/0512138}{{\tt hep-th/0512138}}. 
%
\bibitem{Alishahiha} M. Alishahiha and H. Ebrahim,
``Non-Supersymmetric Attractors and Entropy Function'', 
{\tt hep-th/0601016}.
%
\bibitem{Behrndt:1996jn} K. Behrndt, G.L. Cardoso, B. de Wit,
R. Kallosh, D. L\"ust and T. Mohaupt,
``Classical and quantum {$N=2$} supersymmetric black holes'',
{\em Nucl. Phys.} {\bf B488} (1997) 236-260,
 {\tt hep-th/9610105}.
%
\bibitem{deWit:1996ix} B. de Wit, ``{$N=2$} Electric-magnetic
duality in a chiral background'', 
{\em Nucl. Phys. Proc. Suppl.} {\bf 49}
(1996) 191-200, {\tt hep-th/9602060};
``{$N=2$} symplectic reparametrizations in a chiral background'',
{\em Fortschr. Phys.} {\bf 44} (1996) 529-538,
 {\tt hep-th/9603191}.
%
\bibitem{Sen:1994eb} A.~Sen, ``Black hole solutions in heterotic string
  theory on a torus'', {\it Nucl. Phys.} {\bf B440} (1995) 421,
  \href{http://www.arXiv.org/abs/hep-th/9411187}{{\tt hep-th/9411187}}.
%
\bibitem{Sen:1995in} A.~Sen, ``Extremal black holes and elementary
  string states'', {\em Mod. Phys.  Lett.} {\bf A10} (1995) 2081,
  \href{http://www.arXiv.org/abs/hep-th/9504147}{{\tt
      hep-th/9504147}}.
%
\bibitem{Dabholkar:1989jt} 
A.~Dabholkar and J.A.~Harvey, ``Nonrenormalization of
  the superstring tension'', {\em Phys. Rev. Lett.} {\bf 63} (1989) 478.
%
\bibitem{Cardoso:2004xf} G.L. Cardoso, B. de Wit, J. K\"appeli and T. Mohaupt,
  ``Asymptotic degeneracy of dyonic $N=4$ string states and black hole
  entropy'', {\it JHEP} {\bf 12} (2004) 075, {\tt hep-th/0412287}.
%
\bibitem{LopesCardoso:1999ur}
G.L. Cardoso, B.~de~Wit and T.~Mohaupt, ``Macroscopic entropy formulae and
  non-holomorphic corrections for supersymmetric black holes'', {\em Nucl.
  Phys.} {\bf B567} (2000) 87--110,
\href{http://www.arXiv.org/abs/hep-th/9906094}{{\tt hep-th/9906094}}.
%
\bibitem{Witten:1993ed}
E.~Witten, ``Quantum background independence in string theory'',
{\tt hep-th/9306122}.
%
\bibitem{Freed:1997dp}
D.~S.~Freed, ``Special K\"ahler manifolds'', {\em Comm. Math. Phys.}
{\bf 133} (1999) 31--52, {\tt hep-th/9712042} \,.
%
\bibitem{Alekseevsky} D.V. Alekseevky, V. Cort\'es and C. Devchand,
``Special complex manifolds'', {\em Jour. Geom. Phys.} {\bf 42} (2002) 85,
{\tt math.DG/9910091}.
%
\bibitem{Hitchin} N.J. Hitchin, ``The moduli space of complex
Lagrangian submanifolds'', {\em Asian. J. Math.} {\bf 3} (1999) 77, {\tt
math.DG/9901069}. 
%
\bibitem{Cortes:2001qd}
V.~Cort\'es, ``A holomorphic representation formula for parabolic
hyperspheres'', in: B.~Opozda, U.~Simon and M.~Wiehe (eds.),
{\em PDEs, Submanifolds and Affine Differential Geometry}, Banach
Center Publications, Warsaw (2000), {\tt math.dg/0107037}.
%
\bibitem{Dabholkar:2004dq}
A.~Dabholkar, R.~Kallosh and A.~Maloney, ``A stringy cloak for a classical
  singularity'', {\em JHEP} {\bf 0412} (2004) 059,
\href{http://www.arXiv.org/abs/hep-th/0410076}{{\tt hep-th/0410076}}.
%
\bibitem{Sen:2004dp}
A.~Sen, ``How does a fundamental string stretch its horizon?'',
{\em JHEP} {\bf 0505} (2005) 059,
\href{http://www.arXiv.org/abs/hep-th/0411255}{{\tt hep-th/0411255}}.
%
\bibitem{Hubeny:2004ji}
V.~Hubeny, A.~Maloney and M.~Rangamani, ``String-corrected black holes'',
{\em JHEP} {\bf 0505} (2005) 035,
\href{http://www.arXiv.org/abs/hep-th/0411272}{{\tt hep-th/0411272}}.
%
\bibitem{Chaudhuri:1995fk} 
S. Chaudhuri, G. Hockney and J.D. Lykken, ``Maximally
supersymmetric string theories in $D<10$'', {\em Phys. Rev. Lett.} {\bf 75}
(1995) 2264, {\tt hep-th/9505054}.
%

%
\bibitem{Sen:2005pu} A. Sen, ``Black holes,
elementary strings and holomorphic anomaly'', {\em JHEP} {\bf 0507} (2005) 063,
{\tt hep-th/0502126}.
%

%
\bibitem{Jatkar:2005bh} D.P. Jatkar and A. Sen,
``Dyon spectrum in CHL models'',
{\tt hep-th/0510147}.
%
\bibitem{Harvey:1996ir}
J.A. Harvey and G.W. Moore, ``Fivebrane instantons and {$R^2$} couplings in
  {$N = 4$} string theory'', {\em Phys. Rev.} {\bf D57} (1998) 2323--2328,
\href{http://www.arXiv.org/abs/hep-th/9610237}{{\tt hep-th/9610237}}.
%
\bibitem{Gregori:1996} A. Gregori, E. Kiritsis, C. Kounnas,
N.A. Obers, P.M. Petropoulos and B. Pioline, ``$R^2$ Corrections and
non-perturbative dualities of $N=4$ string ground states'', {\it Nucl.
Phys.} {\bf B510} (1998) 423, {\tt hep-th/9708062}.  
%
\bibitem{Shih:2005he}  D. Shih and X. Yin, ``Exact black hole degeneracies
and the topological string'', {\tt hep-th/0508174}.


\bibitem{Parvizi:2005aa}
  S.~Parvizi and A.~Tavanfar,
  ``Minimal redefinition of the OSV ensemble'', 
{\tt hep-th/0508231}.
%

%
\bibitem{Dijkgraaf:1996it}
R.~Dijkgraaf, E.~Verlinde and H.~Verlinde, ``Counting dyons in $N = 4$ string
  theory'', {\em Nucl. Phys.} {\bf B484} (1997) 543--561,
\href{http://www.arXiv.org/abs/hep-th/9607026}{{\tt hep-th/9607026}}.
%
\bibitem{Shih:2005uc}  S. Shih, A. Strominger and X. Yin, ``Recounting
dyons in {$N=4$} string theory'', {\tt hep-th/0505094}.
%
\bibitem{Ceresole} A. Ceresole, R. D'Auria, S. Ferrara and A. Van
Proeyen, ``Duality transformations in supersymmetric Yang-Mills
theories coupled to supergravity'', {\it Nucl. Phys.} {\bf B444}
(1995) 92, {\tt hep-th/9502072}.  
%
\bibitem{DKLL} B. de Wit, V. Kaplunovsky, J. Louis and D. L\"ust,
``Perturbative couplings of vector multiplets in $N=2$ heterotic
string vacua'', {\it Nucl. Phys.} {\bf B451} (1995) 53, {\tt
hep-th/9504006}.  





\end{thebibliography}
\providecommand{\href}[2]{#2}
\begingroup\raggedright\endgroup
\end{document}